\documentclass[11pt]{article}

\usepackage[final]{acl}

\usepackage{times}
\usepackage{latexsym}
\usepackage{comment}
\usepackage{amsmath}
\usepackage{amsfonts}
\usepackage{graphicx}
\usepackage{booktabs}
\usepackage{tabularx}
\usepackage[T1]{fontenc}
\usepackage[utf8]{inputenc}

\usepackage{tcolorbox}
\usepackage{amssymb}
\usepackage{xcolor}
\usepackage{enumitem}
\usepackage{upquote}
\usepackage{microtype}
\usepackage{multirow}
\usepackage{listings}
\usepackage{inconsolata}
\usepackage{xspace}
\usepackage{courier}
\usepackage{multicol}
\usepackage{url}
\usepackage{hyperref}

\tcbuselibrary{skins, breakable}

\newtcolorbox{promptbox}[1][]{
  colback=gray!10,
  colframe=black!75,
  coltitle=white,
  fonttitle=\bfseries\large,
  fontupper=\ttfamily\small,
  sharp corners,
  boxrule=1.5mm,
  title={#1}
}

\newcommand{\code}[1]{\texttt{#1}}

\newcommand{\approach}{\textit{ExecVerify}\xspace}

\title{\approach: White-Box RL with Verifiable Stepwise Rewards for Code Execution Reasoning}

\author{
\begin{tabular}{c}
{\large\bfseries
Lingxiao Tang$^{1,3}$ \quad
He Ye$^{2}$ \quad
Zhaoyang Chu$^{2}$ \quad
Muyang Ye$^{1}$
} \\[2pt]
{\large\bfseries
Zhongxin Liu$^{1}$ \quad
Xiaoxue Ren$^{1}$ \quad
Lingfeng Bao$^{1,3,*}$
} \\[4pt]
{\normalsize\normalfont
$^{1}$The State Key Laboratory of Blockchain and Data Security, Zhejiang University
} \\
{\normalsize\normalfont
$^{2}$University College London
} \\[3pt]
{\small\normalfont\ttfamily
\{lingxiaotang, yemuyang, liu\_zx, xxren, lingfengbao\}@zju.edu.cn
} \\
{\small\normalfont\ttfamily
he.ye@ucl.ac.uk \quad zhaoyang.chu.25@ucl.ac.uk
}
\end{tabular}
}

\begin{document}
\maketitle

\begingroup
\renewcommand\thefootnote{}
\footnotetext{\textsuperscript{3} Also with Hangzhou High-Tech Zone (Binjiang) Institute of Blockchain and Data Security.}
\footnotetext{\textsuperscript{*} Corresponding author.}
\endgroup

\begin{abstract}
Code LLMs still struggle with code execution reasoning, especially in smaller models. Existing methods rely on supervised fine-tuning (SFT) with teacher-generated explanations, primarily in two forms: (1) input--output (I/O) prediction chains and (2) natural-language descriptions of execution traces. However, intermediate execution steps cannot be explicitly verified during SFT, so the training objective can be reduced to merely matching teacher explanations. Moreover, training data is typically collected without explicit control over task difficulty. We introduce \approach, which goes beyond text imitation by incorporating verifiable white-box rewards derived from execution traces, including next-statement prediction and variable value/type prediction. Our work first builds a dataset with multiple difficulty levels via constraint-based program synthesis. Then, we apply reinforcement learning (RL) to reward correct answers about both intermediate execution steps and final outputs, aligning the training objective with semantic correctness at each execution step. Finally, we adopt a two-stage training pipeline that first enhances execution reasoning and then transfers to code generation. Experiments demonstrate that a 7B model trained with \approach achieves performance comparable to 32B models on code reasoning benchmarks and improves pass@1 by up to 5.9\% on code generation tasks over strong post-training baselines\footnote{We have released our code, data, and models at \url{https://github.com/tlx000000001/ExecVerify}}.
\end{abstract}

\section{Introduction} \label{sec:introduction}
\begin{figure}[t]
  \centering
  \includegraphics[width=\linewidth]{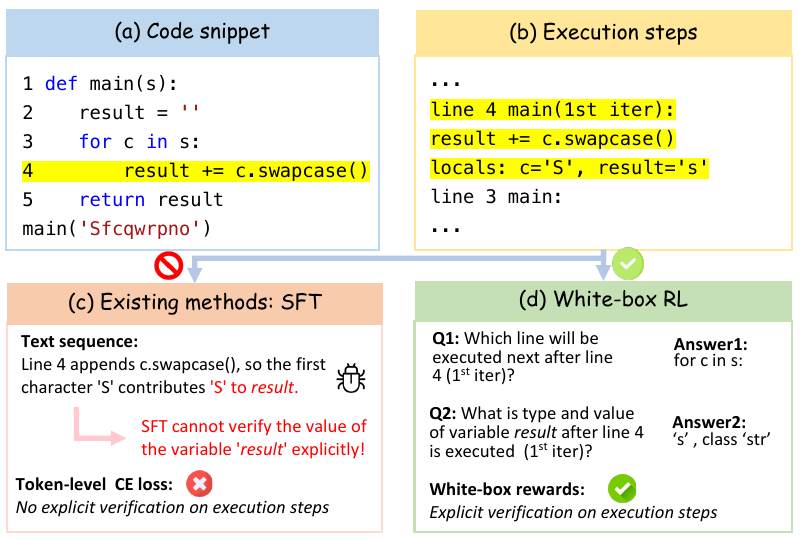}
  \caption{Comparison between SFT and white-box RL. (a) Code snippet. (b) Execution steps extracted from the interpreter, with the relevant parts highlighted in yellow. (c) SFT optimizes the cross-entropy loss over the entire sequence, without explicitly verifying execution details like variable values or control flow. (d) In contrast, white-box RL leverages interpreter-provided execution steps to assign verifiable and step-level rewards.}
  \label{fig:introduction}
\end{figure}

Recent advances in large language models (LLMs)~\cite{hui2024qwen2,zhu2024deepseek} have achieved strong performance on multiple programming tasks~\cite{jiang2024survey,liu2023your,husein2025large}. However, these models often struggle to reason about the concrete execution process of programs~\cite{gu2024cruxeval}. 
This limitation hinders semantic understanding and degrades downstream performance on code generation~\cite{gu2024counterfeit} and program repair~\cite{ni2024next,ye2022neural}. 
A key reason is that the training data is predominantly static text (e.g., source code and docstrings)~\cite{luo2023wizardcoder,kocetkov2022stack}.

% To bridge this gap, prior work incorporates execution information into training, which can be broadly grouped into two lines. 
% I/O-centric approaches (e.g., SEMCODER~\cite{ding2024semcoder} and CODEI/O~\cite{li2025codei}) treat programs as input--output mappings and use execution primarily to validate teacher-generated I/O reasoning chains. 
% Trace-centric approaches (e.g., TracePile~\cite{chen2025chain} and Code Execution as Grounded Supervision for LLM Reasoning~\cite{jung2025code}) leverage full execution traces to construct step-by-step explanations, more directly encouraging models to follow the program's execution process. 
% However, both lines still rely on supervised fine-tuning (SFT) over teacher-produced text: under a token-level cross-entropy objective, models may learn to match the surface form of the teacher’s explanations even when intermediate branches or variable states deviate from the true execution; moreover, SFT itself may generalize poorly~\cite{gupta2025selective,wang2022two}. 
% In addition, training data is often mined or generated passively with limited control over difficulty, resulting in many instances that are either trivial or unsolvable for the target model and lacking an effective curriculum (Appendix~\ref{subsection:Appendix-Difficulty-Imbalance}).

To bridge this gap, prior work has incorporated execution signals into training, primarily through two approaches: I/O-centric methods (e.g., SEMCODER~\cite{ding2024semcoder}, CODEI/O~\cite{li2025codei}), which use execution to validate teacher-generated input–output reasoning chains, and trace-centric methods (e.g., TracePile~\cite{chen2025chain}, Code Execution as Grounded Supervision~\cite{jung2025code}), which convert execution traces into step-by-step explanations.
However, both approaches typically rely on SFT over teacher-written text. Under a token-level cross-entropy objective, intermediate execution steps are not explicitly verified during training. Figure~\ref{fig:introduction} illustrates this limitation of SFT and compares it with our white-box RL approach. As a result, models may overfit to the teacher’s textual explanations without truly understanding the execution process. Furthermore, SFT has shown limited generalization ability~\cite{gupta2025selective,wang2022two}.
In addition, training data is often passively collected or generated without control over difficulty, resulting in many examples that are either trivial or unsolvable, and lacking a structured learning curriculum (see Appendix~\ref{subsection:Appendix-Difficulty-Imbalance}). 

We introduce \approach, a framework that enhances execution reasoning by combining Constraint-Based Data Synthesis and White-Box Reinforcement Learning.
First, we synthesize programs under explicit structural constraints to construct a curriculum-style dataset with multiple difficulty levels, covering a broad range of commonly used data types and built-in methods.
Next, as shown in Figure~\ref{fig:introduction}, we convert interpreter traces into verifiable white-box questions that target intermediate control flow, as well as variable types and values. We then apply reinforcement learning (RL) to reward the model for correct predictions on both intermediate steps and final outputs, shifting the objective from text-level imitation to semantic understanding of the execution process.
Finally, we adopt a two-stage post-training strategy: the first stage strengthens execution reasoning through white-box rewards, and the second adapts the model to code generation using unit-test feedback, enabling effective transfer from reasoning to generation.

Extensive experiments demonstrate the effectiveness of \approach. 
On execution reasoning benchmarks, a 7B model trained with \approach achieves strong results on CRUXEval~\cite{gu2024cruxeval}, LiveCodeBench-Exec~\cite{jain2024livecodebench}, and REval~\cite{chen2024reasoning}, and is competitive with much larger models such as Qwen2.5-Coder-32B-Instruct~\cite{hui2024qwen2}. 
Building on this foundation model, when further post-trained for code generation, our model consistently outperforms strong post-training baselines on mainstream benchmarks, including EvalPlus~\cite{evalplus}, LiveCodeBench~\cite{jain2024livecodebench}, and BigCodeBench~\cite{zhuo2024bigcodebench}, yielding up to a 5.9\%  improvement in pass@1.

\section{\approach}
We propose \approach, as shown in Figure~\ref{fig:codereasoner-overview}, which improves the LLM's ability in code execution reasoning via \textbf{Constraint-Based Data Synthesis (upper part)} and \textbf{Two-Stage Post-Training (bottom part)}. \approach first synthesizes programs with controlled difficulty under structural constraints. It then applies the Two-Stage Post-Training pipeline. Step one uses verifiable white-box rewards from execution traces for code execution reasoning and step two utilizes unit-test rewards for code generation.

\subsection{Constraint-Based Data Synthesis}
\label{sec:data_synthesis}
\begin{figure*}[t]
    \centering
    % 建议把图片文件放到 figures/ 目录下并改个简短名字
    \includegraphics[width=1.0\linewidth]{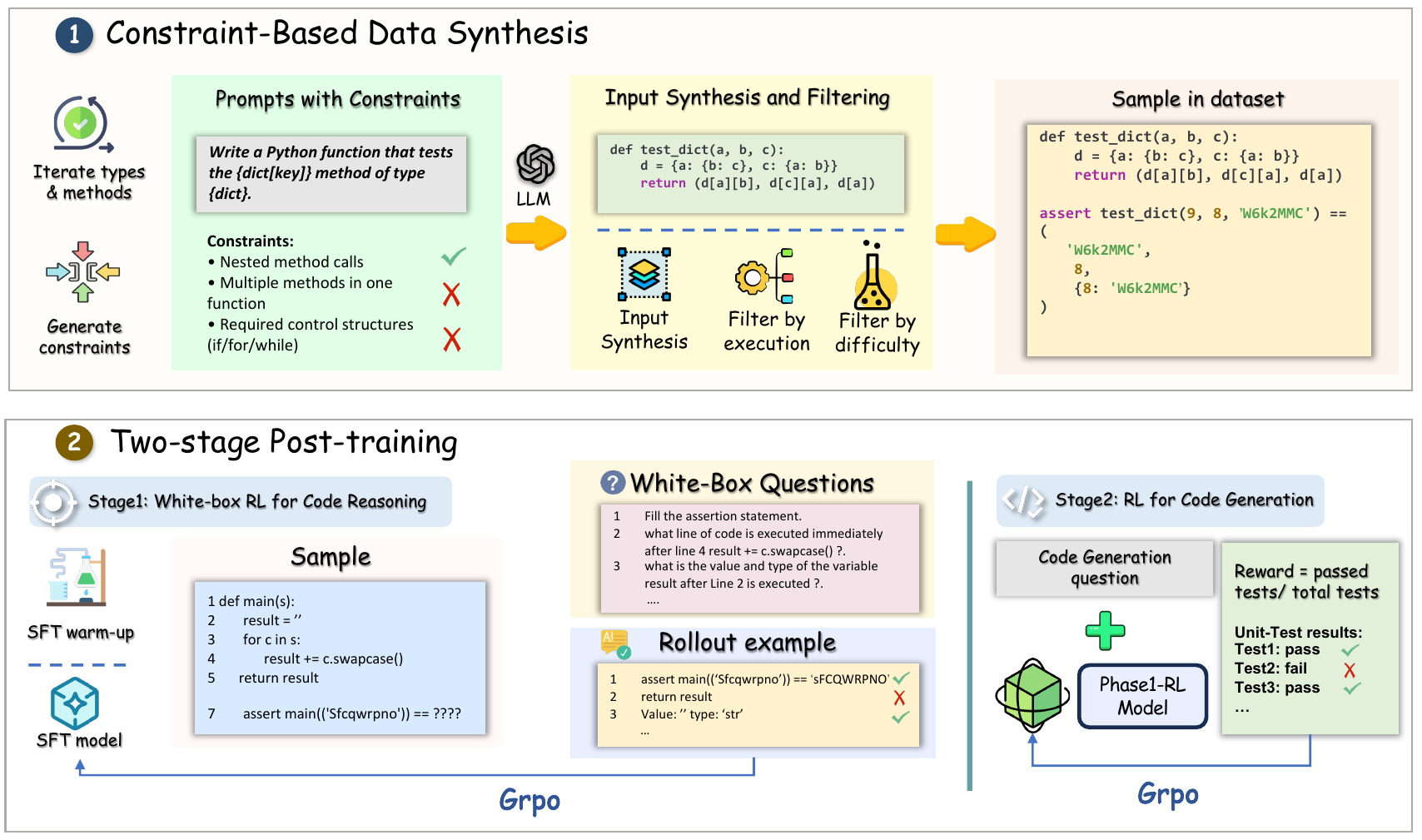}
    \caption{Overview of our approach. 
    Step~1 constructs a constraint-based dataset of executable Python snippets. 
    Step~2 performs two-stage post-training: white-box RL for code reasoning
    followed by RL for code generation.}
    \label{fig:codereasoner-overview}
\end{figure*}

The goal of Constraint-Based Data Synthesis is twofold: (i) to ensure structural diversity by systematically covering common types, methods, and control-flow patterns; and (ii) to ensure controlled difficulty by generating programs across multiple difficulty levels that remain challenging yet solvable for smaller models. This contrasts with prior methods that collect data without control over structure and difficulty. This component corresponds to the upper part of Figure~\ref{fig:codereasoner-overview}.
% We introduce our work CodeReasoner which actively synthesize executable Python programs under explicit structural constraints, contrast to/different from the prior work that passively produces data without fine-grained control.
% In our approach emphasize the executability, difficulty and diversity of the generated inputs and filter based on   constraint-based data synthesis, illustrated in the upper part in Figure~\ref{fig:codereasoner-overview}.

\subsubsection{Prompt with Constraints}
\paragraph{Iterating types and methods.}
\approach begins by iterating over all built-in Python types and their associated methods. For each method, the LLM is explicitly asked to generate a piece of code that must contain the mentioned type and the method.

\paragraph{Generating constraints.}
To increase complexity, we incrementally apply two types of structural constraints during prompting:
\textbf{(i) Method-call constraints}, which require the LLM to use nested calls and combine multiple methods within a single function, encouraging rich method interactions; and
\textbf{(ii) Control-structure constraints}, which enforce the presence of specific nested control-flow patterns, such as \texttt{while}, \texttt{for}, or \texttt{if} statements, to produce non-trivial execution paths (see Appendix~\ref{app:code-synthesis-prompts} for examples). 

\paragraph{From simplicity to complexity.}
These constraints are introduced in stages: starting with simple code that uses a single method, we then apply method-call constraints, and finally add control structures with increasing nesting depth. This process produces programs that evolve naturally from simple to complex.

\subsubsection{Input Synthesis and Data Filtering} 
\label{subsec:Input-Synthesis-and-Data-Filtering}
\paragraph{Input synthesis.} To probe program behavior, we generate diverse inputs for each code snippet. We first prompt an LLM to produce an initial input (as an assertion on the entry-point function), and then apply type-aware mutation following Liu et al.~\cite{liu2023your} to obtain additional valid inputs. This yields multiple executable inputs per snippet, including both original and mutated variants. In total, we generate \textbf{239,992} raw and \textbf{239,466} mutated instances before filtering (see Appendix~\ref{subsection:appendix-filtering-statistics} and Appendix~\ref{subsection:Appendix-Input-Synthesis-Mutation}).
\begin{comment}
To probe program behavior and reduce overfitting to a single input configuration, we generate diverse inputs for each synthesized code snippet.
We first prompt an LLM to generate an initial input, typically in the form of an assertion statement on the entry-point function. Then, following Liu et al.~\cite{liu2023your}, we perform type-aware mutation by providing the LLM with the original code, a pool of candidate integers and strings, and mutation guidelines. We encourage the LLM to generate arguments with longer strings or containers and using reference values from the pool. This process yields multiple valid inputs for each code snippet, including both original and mutated variants. In total, we generate 239,992 raw and 239,466 mutated instances before filtering (see Appendix~\ref{subsection:appendix-filtering-statistics} for  statistics and Appendix~\ref{subsection:Appendix-Input-Synthesis-Mutation} for examples).
\end{comment}

\paragraph{Filtering by execution.} We execute each synthesized program on all its candidate inputs and discard instances that fail to run successfully or violate basic output constraints (e.g., runtime exceptions, timeouts, or excessively long outputs).
As a result, this filtering process leads to \textbf{201,537} raw and \textbf{191,463} mutated instances.

\paragraph{Filtering by difficulty.} \approach encourages our model to learn from challenging data points. To filter out trivial samples, we evaluate each remaining instance using Qwen2.5-Coder-7B-Instruct~\cite{hui2024qwen2} under the input–output prediction setting. Specifically, we run the model ten times at temperature 1.0 and count how many predictions pass the test cases. We retain only instances with at most three successful runs (pass count $\leq 3$), resulting in \textbf{119{,}358} training examples in total. This yields instances that are non-trivial yet solvable for small models. We report the resulting difficulty and complexity distributions in Appendix~\ref{subsection:Appendix-Difficulty-Complexity-Distribution}--\ref{subsection:appendix-filtering-statistics}.

\paragraph{Contamination analysis.}
We also perform an embedding-based contamination analysis against all test sets and find no instances exceeding a conservative similarity threshold (see Appendix~\ref{subsection:Appendix-Contamination-Analysis}).

\subsection{Two-stage Post-training}
\label{sec:two_stage_post_training}

As shown in the bottom part of Figure~\ref{fig:codereasoner-overview}, our training pipeline consists of two stages:
Stage I enhances execution reasoning using white-box rewards, while Stage II adapts the model to code generation through unit-test feedback.
% 把2.2.1和2.2.2塞一起

\subsubsection{Stage I: White-box RL for Code Reasoning} % 强调一下为什么
The goal of Stage~I is to strengthen the execution reasoning ability by training the model to predict both intermediate execution states and final outputs. This shifts learning from the prior paradigm of imitating teacher explanations to stepwise semantic correctness during execution. 

Our work starts with a brief warm-up to inject execution-aware reasoning patterns. Specifically, we apply supervised fine-tuning (SFT) on input-output prediction reasoning chains generated by a strong teacher model~\cite{team2024qwq} and filtered via rejection sampling to ensure correctness. This warm-up provides the model with fundamental execution-relevant reasoning behaviors, which are difficult to discover through reinforcement learning alone~\cite{yue2025does}.

We then switch to reinforcement learning with output correctness as the reward. However, the I/O-based rewards only evaluate the final output and fail to assess intermediate execution steps. To overcome this, we introduce \emph{white-box reward signals}, which generate verifiable questions from execution traces and reward the model based on its predictions of control flow and variable states, including values and types.

\paragraph{Trace collection from the interpreter.}

Given a synthesized program $f$ and input $x$, we execute $f(x)$ using an interpreter to obtain an execution trace $\tau = {(l_t, \sigma_t)}_{t=1}^{T}$, where $l_t$ is the executed statement at step $t$, and $\sigma_t$ is the program state, including values and types of in-scope variables.

\paragraph{White-box question construction.}
From the execution trace $\tau$, we deterministically construct two types of white-box questions:
(i) \emph{Control-flow questions}, which ask the model to predict the next executed statement $l_{t+1}$;
(ii) \emph{Data-flow questions}, which ask the model to predict updated variable values and types in $\sigma_{t+1}$.
All questions are generated automatically from the interpreter and have a unique verifiable answer derived from the trace (see Appendix~\ref{appendix:white-box_question_generation} for details).

\paragraph{White-box reward function.}
We design the white-box reward function as follows:
\[
R_{\text{white-box}}
= 2 \cdot \Big( (1-\alpha)\, R^{(\text{I}\rightarrow\text{O})}
+ \alpha\, R_{\text{white}} \Big),
\]
where $\alpha \in [0,1]$ balances the weight of final I/O correctness and white-box execution accuracy, and the factor of 2 ensures that the overall reward value ranges from 0 to 2.
The term $R^{(\text{I}\rightarrow\text{O})}$ measures correctness under the input--output prediction setting and is a binary reward that takes value 1 if the model’s predicted output matches the ground-truth output, and 0 otherwise.
The term $R_{\text{white}}$ measures the model’s accuracy in predicting intermediate execution states and is computed over a sampled set $Q_s$ of white-box questions:
\[
R_{\text{white}} = \frac{1}{|Q_s|} \sum_{q_j \in Q_s} \mathbb{I}[a_j = a_j^*],
\]
where $a_j$ is the model’s answer to question $q_j$, and $a_j^*$ is the corresponding ground-truth answer derived from execution traces. This term reflects the model’s accuracy in predicting intermediate execution states.

\paragraph{O$\rightarrow$I prediction reward function.} % 加一个目的
To encourage the model to reason in both directions and reduce reliance on forward input-to-output pattern matching, we also include reverse prediction tasks where the model predicts inputs from outputs. Since one output may have multiple valid inputs, we do not define white-box questions in this case. Instead, we assign a reward of 2 if the predicted input produces the correct output when executed, and 0 otherwise, maintaining the same $[0,2]$ reward scale for consistency.

\subsubsection{Stage II: RL for code generation} % 把这一段稍微写长一点，重要一点
\label{subsec:RL_for_code_generation}
Once the model has obtained the code reasoning ability in the first stage, we further post-train it for the code generation task. The goal of this stage is to align the model’s execution reasoning abilities with the objective of generating functionally correct programs. Following the previous study~\cite{cui2025process}, we use a reward $R^{(gen)}$ defined as the proportion of unit tests the generated solution successfully passes:

$$R^{(gen)} = \frac{\text{Number of passed tests}}{\text{Total number of tests}}$$

This reward signal guides the model to apply its execution reasoning capabilities to generate functionally correct code.

\section{Experiment Setup}
\label{sec:experiment_setup}
\subsection{Training Details} % 把Table 1的variant的数据细节提到这里来，training的参数全放附录
We use Qwen2.5-Coder-7B-Instruct~\cite{hui2024qwen2} as the base model. We perform full-parameter SFT for the warm-up stage, and then apply GRPO~\cite{guo2025deepseek} for both Stage~I and Stage~II. We use a maximum sequence length of 4096 for training, and for RL we sample $n=8$ rollouts per prompt for 500 steps with a KL coefficient of 0.0. Full SFT and RL configurations are provided in Appendix~\ref{subsection: Appendix-Supervised-Fine-Tuning-(SFT)} and Appendix~\ref{subsection: Reinforcement-Learning-(GRPO)}.
For Stage~I, we use 30K synthesized samples for the SFT warm-up and another 30K for white-box RL. For each RL instance, we sample up to 10 white-box questions to compute $R_{\text{white}}$ (see full setup in Appendix~\ref{app:variant-setup}). We set $\alpha=0.5$ to balance terminal I/O reward and step-level white-box accuracy. Varying $\alpha$ in $\{0.25, 0.5, 0.75\}$ yields similar performance (see Appendix~\ref{appendix：coefficient-sensitivity}).
For code-generation RL, we use the PrimeCode~\cite{cui2025process} dataset, sourced from APPS~\cite{hendrycks2021measuring}, CodeContests~\cite{li2022competition}, TACO~\cite{li2023taco}, and CodeForces~\cite{penedo2025codeforces}.

\subsection{Benchmarks}

For code reasoning, we evaluate on three widely used benchmarks: 
CRUXEval~\cite{gu2024cruxeval}, 
LiveCodeBench-Exec~\cite{jain2024livecodebench}, 
and REval~\cite{chen2024reasoning}, following the settings of prior work~\cite{li2025codei,ding2024semcoder,chen2025chain}. 
The REval benchmark evaluates whether the model can correctly infer control flow, variable values, and variable types during the execution process.
For code generation, we evaluate on three standard benchmarks: 
EvalPlus~\cite{evalplus}, 
LiveCodeBench-V6~\cite{jain2024livecodebench}, 
and BigCodeBench~\cite{zhuo2024bigcodebench}. 
All evaluations use greedy sampling with the temperature set to 0.0, and we report pass@1 as the evaluation metric.

\subsection{Baselines}

For code reasoning, we compare our model against strong large-sized LLMs, including 
Qwen2.5-Coder-32B-Instruct~\cite{hui2024qwen2} 
and Llama3-Instruct-70B~\cite{dubey2024llama}. We additionally include SEMCODER~\cite{ding2024semcoder} and CODEI/O~\cite{li2025codei}, both tuned on Qwen2.5-Coder-7B-Instruct. For code generation, we include larger models like 
Llama3-Instruct-70B~\cite{dubey2024llama}, 
DeepSeek-Coder-V2-Lite-Instruct~\cite{zhu2024deepseek}, 
and Qwen2.5-Coder-14B-Instruct~\cite{hui2024qwen2} as baselines. 
 SEMCODER~\cite{ding2024semcoder} and CODEI/O~\cite{li2025codei} are also evaluated on code generation benchmarks for completeness.

\section{Experimental Results}
\label{sec:experiment_results}
\begin{table*}[t]
\centering
\small
\setlength{\tabcolsep}{4pt}
\caption{Code reasoning experimental results on CRUXEval, LiveCodeBench, and REval. Average is computed over all fine-grained metrics.  }
\begin{tabular}{llcccccccc}
\toprule
Model & Size & \multicolumn{2}{c}{CRUXEval} & \multicolumn{1}{c}{LiveCodeBench-Exec} & \multicolumn{4}{c}{REval} & \multirow{2}{*}{Average} \\
\cmidrule(lr){3-4}\cmidrule(lr){5-5}\cmidrule(lr){6-9}
 &  & CXEval-O & CXEval-I & LCB-O & Coverage & State & Path & Output & \\
\midrule
Qwen2.5-Coder-Instruct         & 32B & 78.2 & 83.4 & 80.6 & 84.6 & 66.7 & 68.7 & 83.3 & 77.9 \\
LLaMA3-Instruct                & 70B & 63.7 & 61.3 & 56.4 & 85.3 & 59.2 & 40.3 & 74.6 & 63.0 \\
SemCoder                       & 7B  & 66.7 & 65.3 & 58.9 & 80.8 & 54.7 & 50.7 & 65.3 & 63.2 \\
CODEI/O                        & 7B  & 62.5 & 67.9 & 60.8 & 70.9 & 52.5 & 45.9 & 60.8 & 60.1 \\
\midrule
Qwen2.5-Coder-Instruct         & 7B  & 61.0 & 66.0 & 58.0 & 78.6 & 51.7 & 49.7 & 60.3 & 60.8 \\
+ I/O O/I RL          & 7B  & 74.9 & 76.5 & 72.2 & 83.2 & 65.1 & 50.4 & 77.3 & 71.4 \\
+ SFT + I/O O/I RL    & 7B  & 83.5 & \textbf{84.3} & 81.0 & 84.4 & 67.2 & 51.4 & 82.3 & 76.3 \\
+ SFT + white-box RL  & 7B  & \textbf{85.6} & 81.0 & \textbf{82.3} & \textbf{85.8} & \textbf{74.5} & \textbf{73.0} & \textbf{83.2} & \textbf{80.8} \\
\bottomrule
\end{tabular}

\label{tab:code_reasoning}
\end{table*}

\subsection{Code Reasoning Results}

\paragraph{Our SFT + white-box RL model outperforms strong baseline models. Both SFT and white-box RL are effective.}
Table~\ref{tab:code_reasoning} summarizes our experimental results on CRUXEval, LiveCodeBench-Exec, and REval. 
Across comparable 7B variants trained on the same synthesized corpus (details in Appendix~\ref{app:variant-setup}), I/O RL substantially improves over our base model Qwen2.5-Coder-7B-Instruct, adding SFT warm-up yields further gains, and replacing  I/O RL with white-box RL achieves the best overall performance.
As shown in the last column, our final model improves the average score from 60.8 to 80.8 (+20.0) and is competitive with Qwen2.5-Coder-32B-Instruct (77.9).

\subsection{Code Generation Results}

\begin{table*}[t]
\centering
\small
\setlength{\tabcolsep}{4pt}
\caption{Code generation results on HumanEval, MBPP, LiveCodeBench, and BigCodeBench. Average is computed over all fine-grained metrics.  }
\resizebox{\linewidth}{!}{%
\begin{tabular}{llcccccccccc}
\toprule
Backbone & Size & \multicolumn{2}{c}{HumanEval} & \multicolumn{2}{c}{MBPP} & \multicolumn{3}{c}{LiveCodeBench} & \multicolumn{2}{c}{BigCodeBench} & \multirow{2}{*}{Average} \\
\cmidrule(lr){3-4}\cmidrule(lr){5-6}\cmidrule(lr){7-9}\cmidrule(lr){10-11}
 &  & HE & HE+ & MBPP & MBPP+ & Easy & Medium & Hard & Full & Hard & \\
\midrule
Qwen2.5-Coder-Instruct        & 32B & {92.7} & {87.2} & {90.2} & {75.1} & {84.5} & {22.0} & 3.4 & 49.6 & {27.0} & {59.0} \\
\midrule
Qwen2.5-Coder-Instruct        & 14B & 89.6 & \textbf{87.2} & 86.2 & 72.8 & 71.0 & 13.5 & 2.5 & 48.4 & 22.2 & 54.8 \\
DS-Coder-V2-Lite-Inst.        & 16B & 81.1 & 75.6 & 82.8 & 70.4 & 61.2 &  9.6 & 3.1 & 36.8 & 16.2 & 48.5 \\
LLaMA3-Instruct               & 70B & 77.4 & 72.0 & 82.3 & 69.0 & 61.0 &  9.6 & 3.2 & \textbf{54.5} & \textbf{27.0} & 50.7 \\
SemCoder                      & 7B  & 88.6 & 83.8 & 85.9 & 71.0 & 61.8 & 12.8 & 2.9 & 42.3 & 19.5 & 52.1 \\
CODEI/O                       & 7B  & 86.0 & 80.5 & 81.0 & 69.4 & 56.5 &  7.2 & 2.9 & 40.5 & 17.2 & 49.0 \\
\midrule
Qwen2.5-Coder-Instruct        & 7B  & 88.4 & 84.1 & 83.5 & 71.7 & 60.0 &  9.2 & 3.0 & 41.0 & 18.2 & 51.0 \\
+ UT RL          & 7B  & 87.2 & 82.3 & 85.3 & 72.1 & 64.4 & 13.5 & 4.1 & 49.3 & 23.0 & 53.9 \\
+ I/O RL + UT RL       & 7B  & 90.2 & 82.9 & 85.2 & 71.2 & 68.2 & {15.6} & 3.9 & \textbf{50.6} & 23.3 & 54.6 \\
+ SFT + I/O RL + UT RL & 7B  & \textbf{91.5} & 84.0 & 86.1 & 72.3 & 69.5 & 14.4 & 3.9 & 49.6 & 23.1 & 54.9 \\
+ SFT + white-box RL+ UT RL  & 7B  & {90.9} & {84.8} & \textbf{89.4} & \textbf{75.1} & \textbf{74.5} & \textbf{18.4} & \textbf{5.9} & 49.4 & \textbf{25.7} & \textbf{57.1} \\
\bottomrule
\end{tabular}
}
\label{tab:code_generation}
\end{table*}

\paragraph{Two-stage RL is effective.}
Table~\ref{tab:code_generation} presents code generation results on HumanEval, MBPP, LiveCodeBench, and BigCodeBench. 
We further train the models in Table~\ref{tab:code_reasoning} with unit-test RL (recall in Section~\ref{subsec:RL_for_code_generation}). 
While single-stage GRPO (UT RL) already improves the base 7B model (53.9), initializing GRPO from a reasoning-enhanced checkpoint yields consistently better performance (see last three rows). 
Our best two-stage variant (SFT + white-box RL + unit-test GRPO) achieves the highest average score (57.1) and improves pass@1 by up to 5.9 points over the pure-GRPO baseline.

\paragraph{Code generation benefits from white-box reinforcement learning.} 
Among models trained with unit-test RL, the progression from UT-RL (53.9) to
I/O RL + UT-RL (54.6), SFT + I/O RL + UT-RL (54.9), and finally
SFT + white-box RL + UT-RL (57.1) shows a consistent upward trend. These results indicate that the fine-grained execution knowledge learned via white-box RL, such as
tracking control flow and variable states, not only boosts code reasoning performance but also
transfers to realistic code generation tasks.

\subsection{Data Efficiency}

\begin{figure}[t]
  \centering
  \includegraphics[width=\linewidth]{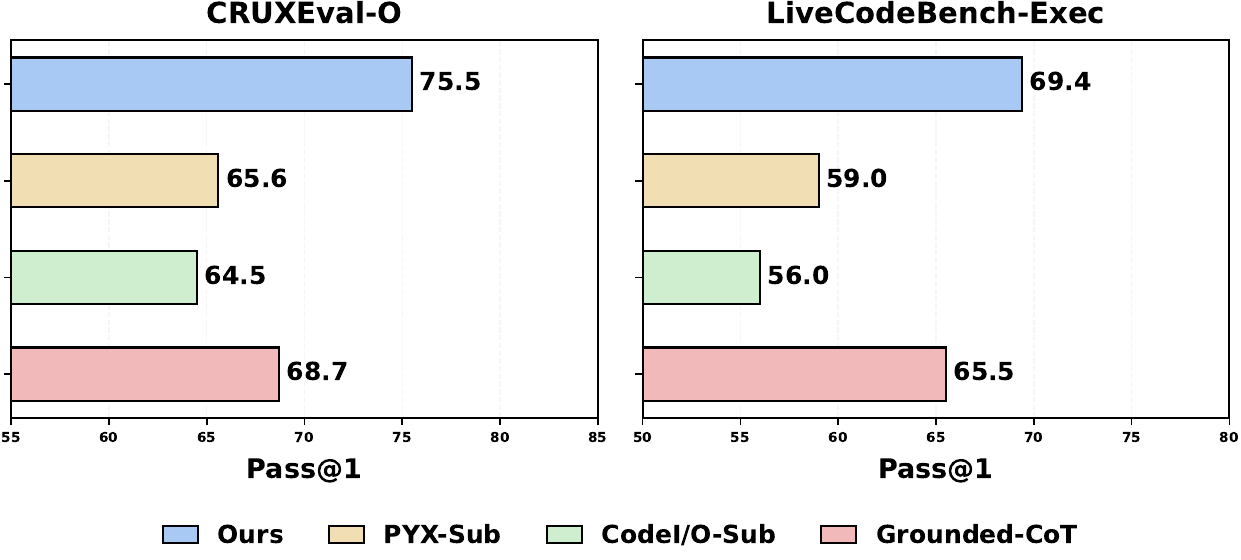}
  \caption{Data efficiency comparison at a fixed training scale (15K examples). We report Pass@1 on CRUXEval-O and LiveCodeBench-Exec for models fine-tuned with different datasets.}
  \label{fig:data-efficiency-scaling}
\end{figure}
% caption放大一点

To isolate the impact of data quality, we compare our synthesized data with three representative datasets: PYX-Sub~\cite{ding2024semcoder}, CodeI/O-Sub~\cite{li2025codei}, and Grounded-CoT~\cite{jung2025code}, which correspond to the two execution-supervision paradigms in Section~\ref{sec:introduction} (I/O CoT vs.\ LLM-translated traces). 
We sample 15K instances randomly from each training dataset (see full setup in Appendix~\ref{app:data-quality-setup}). 
As shown in Figure~\ref{fig:data-efficiency-scaling}, our dataset achieves the highest performance on both CRUXEval-O and LiveCodeBench-Exec, demonstrating superior data efficiency.% 下去改一下语句

\subsection{Ablation on Data Synthesis}
\begin{figure}[t]
  \centering
  \includegraphics[width=\linewidth]{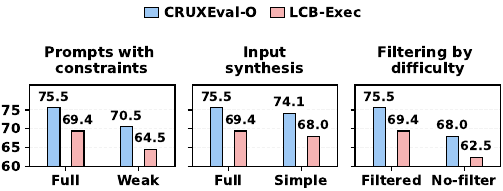}
  \caption{Ablation study on our synthesis pipeline on CRUXEval-O and LiveCodeBench-Exec. We report pass@1 on models finetuned with different data synthesis variants.}
  \label{fig:data_ablation}
\end{figure}

Our synthesis pipeline has three key components: (i) generating prompts with structural constraints, (ii) input synthesis, and (iii) filtering by difficulty. To isolate their contributions, we run three SFT-only ablations on the I/O prediction task using 15K training examples (see Appendix~\ref{app:data-ablation-setup} for all configurations). Figure~\ref{fig:data_ablation} shows that each component improves pass@1 on both benchmarks.

% 在正文适当位置插入

% 正文中

\subsection{Cross-language Generalization}
\begin{figure}[t]
  \centering
  \includegraphics[width=0.98\linewidth]{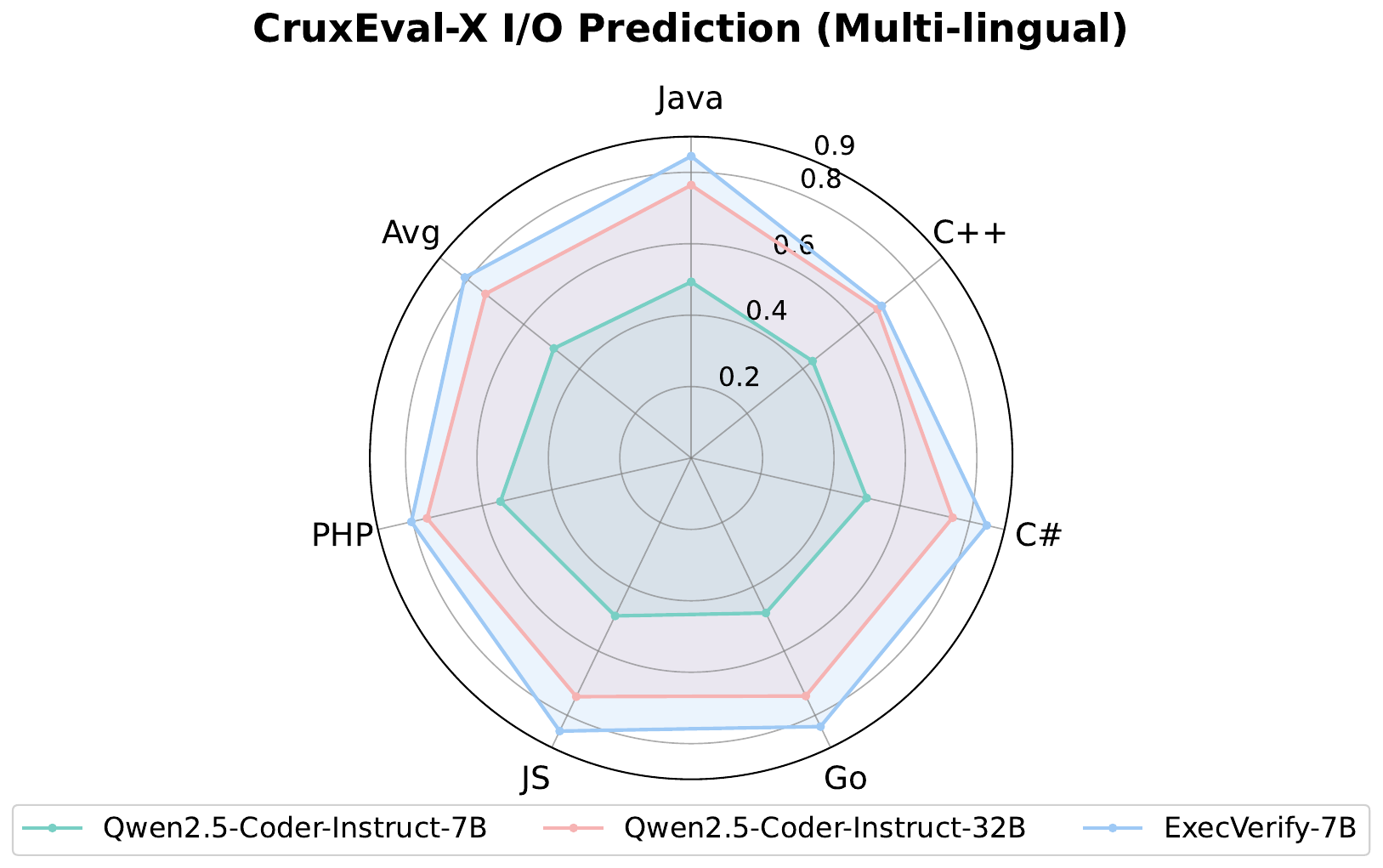}
  \caption{CRUXEval-X Multilingual I/O Prediction: Comparison with Qwen2.5-Coder-Instruct (7B/32B).}
  \label{fig:cross-language-generalization}
\end{figure}

CRUXEval-X~\cite{xu2025cruxeval} is a multilingual code execution reasoning benchmark. To test whether our execution reasoning improvements extend beyond Python, we evaluate I/O prediction on CRUXEval-X across six programming languages (Java, C++, C\#, Go, JavaScript, and PHP). “Avg” denotes the average accuracy across these six languages. As shown in Figure~\ref{fig:cross-language-generalization}, \approach-7B consistently outperforms the same-family base model Qwen2.5-Coder-7B-Instruct across all languages, and is also competitive with the much larger Qwen2.5-Coder-32B-Instruct. These results show that the execution reasoning ability transfers effectively across programming languages.

\subsection{Library-involved I/O Prediction}
\begin{figure}[t]
  \centering
  \includegraphics[width=\linewidth]{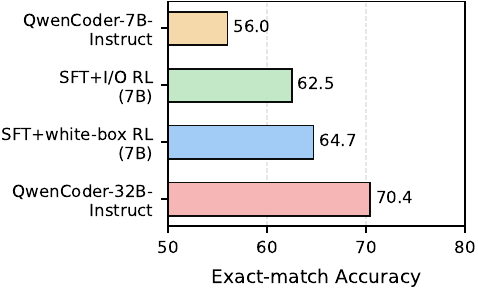}
  \caption{Experimental results on Library-involved I/O Prediction}
  \label{fig:library_involved_io_prediction}
\end{figure}

% To test transfer beyond built-in-only synthetic snippets, we evaluate a more realistic library-involved I/O prediction setting constructed from the I/O examples of BigCodeBench, and report exact-match accuracy on \texttt{stdout}. 
% We parse \texttt{import} statements via Python AST and filter out environment-dependent or non-deterministic behaviors (blacklist in Table~\ref{tab:libio-blacklist}), yielding 241 test cases.
% As shown in Figure~\ref{fig:library_involved_io_prediction}, \textsc{QwenCoder-7B-Instruct} achieves 56.0, \textsc{SFT+I/O RL} 62.5, and \textsc{SFT+white-box RL} 64.7 (vs.\ \textsc{QwenCoder-32B-Instruct} 70.4), indicating that our gains transfer to library-involved settings.

To evaluate \approach on code that depends on external libraries, we construct a library-involved I/O prediction benchmark from BigCodeBench. For each task, we extract an input–output pair from the interactive example in the task description and use the task’s canonical solution as the executable source program. At evaluation time, we provide the extracted input and ask the model to predict the output of the canonical solution. We report the exact match accuracy.  To ensure determinism, we filter out tasks involving randomness or external resources (e.g., random number generation, file I/O), resulting in 241 test cases (see Appendix~\ref{app:library-involved-i/o-prediction-benchmark-construction}). 

As shown in Figure~\ref{fig:library_involved_io_prediction}, {Qwen2.5-Coder-7B-Instruct} achieves 56.0, \textsc{SFT+I/O RL} improves to 62.5, and \textsc{SFT+White-Box RL} further reaches 64.7, compared to 70.4 from the much larger {Qwen2.5-Coder-32B-Instruct}. These results indicate that our improvements transfer to library-involved code settings.

\subsection{Ablations on White-box Questions}
\begin{table*}[t]
\caption{Control-flow vs.\ data-flow ablations for Stage I white-box RL. Best results in each column are highlighted in bold. Average is the mean over all metrics.}
\label{tab:cf-df-ablation}
\centering
\small
\setlength{\tabcolsep}{6pt}
\begin{tabular}{l cc c cccc c}
\toprule
Model & \multicolumn{2}{c}{CRUXEval} & LiveCodeBench & \multicolumn{4}{c}{REval} & Average \\
\cmidrule(lr){2-3}\cmidrule(lr){4-4}\cmidrule(lr){5-8}
 & CXEval-O & CXEval-I & LCB-O & Coverage & State & Path & Output &  \\
\midrule
Full    & \textbf{85.6} & 81.0 & \textbf{82.3} & 85.8 & 74.5 & 73.0 & \textbf{83.2} & \textbf{80.8} \\
CF-only & 84.1 & \textbf{82.6} & 80.9 & \textbf{86.0} & 68.8 & \textbf{74.7} & 82.5 & 79.9 \\
DF-only & 84.9 & 82.1 & \textbf{82.3} & 84.9 & \textbf{75.7} & 54.6 & 82.1 & 78.1 \\
\bottomrule
\end{tabular}
\end{table*}

We ablate the two types of white-box questions by scoring either only control-flow questions (CF-only) or only data-flow questions (DF-only), while keeping the prompting format unchanged (i.e., unscored questions are still generated).
Table~\ref{tab:cf-df-ablation} shows that the two signals are complementary: CF-only improves control-flow metrics but reduces state accuracy, while DF-only enhances variable state prediction but hurts CF performance.
Scoring both types (Full) yields the best overall balance and the highest average score.

\subsection{Training Dynamics} % 图在figure 15,具体分析在D1    
We additionally report training dynamics in Figure~\ref{fig:training-dynamics}. Stage~I white-box RL provides stable improvements in both white-box accuracy and I/O prediction. Stage~II consistently outperforms training the base model from scratch on code generation across the entire training process. Detailed analysis is provided in Appendix~\ref{appendix:training_dynamics}.

\subsection{Generalization across Model Sizes and Architectures}

\begin{figure}[t]
  \centering
  \includegraphics[width=0.9\linewidth]{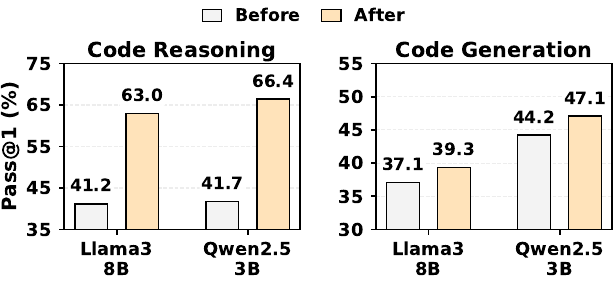}
  \caption{Averaged performance metrics on Code Reasoning and Generation benchmarks. The results demonstrate the generalization of our method across various model sizes and architectures.}
  \label{fig:other-models-avg}
\end{figure}
We further assess the generalizability of our approach across model sizes and architectures by applying the same training pipeline to Qwen2.5-Coder-3B-Instruct~\cite{hui2024qwen2} and Llama3-8B-Instruct~\cite{dubey2024llama}. As shown in Figure~\ref{fig:other-models-avg}, our method yields consistent improvements on Code Reasoning and Code Generation evaluations, indicating that the proposed pipeline transfers beyond a single model family and supports robust gains across different LLM architectures.

\section{Related Work}
\label{sec:related_work}
\paragraph{Enhance LLM's Performance on Code Execution Reasoning} 

Previous works~\cite{liu2023code,ding2024traced} fine-tune LLMs such as UnixCoder~\cite{guo2022unixcoder} directly on raw execution traces. 
Self-Debugging~\cite{chen2023teaching} further finds that directly feeding raw traces can even undermine LLM's performance on program repair. 
To alleviate this, Next~\cite{ni2024next} injects execution information into debugging comments when fine-tuning the model. 
More recent works adopt other training paradigms. 
SemCoder~\cite{ding2024semcoder} and CODEIO~\cite{li2025codei} fine-tune models on input–output and output–input reasoning chains extracted from stronger teacher models. 
TracePile~\cite{chen2025chain} and Code Execution as Grounded Supervision~\cite{jung2025code} also rely on a teacher LLM to translate execution traces into natural-language and perform supervised fine-tuning on it.
In contrast, \approach adopts a new training paradigm: it converts execution traces into white-box questions about control flow and data flow. Therefore, it provides dense and verifiable rewards for intermediate execution steps throughout reinforcement learning.

\paragraph{Evaluating LLMs on Code Execution Reasoning}
Early works~\cite{liu2023code,ding2024traced} evaluate trained models on their own collected datasets. 
More recently, CRUXEval~\cite{gu2024cruxeval} was proposed as a public benchmark for code execution reasoning, and CRUXEval-X~\cite{xu2025cruxeval} extends it to multiple programming languages. 
REval~\cite{chen2024reasoning} further refines the prediction task by requiring LLMs to predict intermediate execution states rather than only final outputs and 
CORE~\cite{xie2025core} evaluates LLMs on more complex static analysis tasks.

\paragraph{Data Synthesis for Code LLMs}
Researchers have proposed multiple synthesized datasets for training code generation models, such as CodeAlpaca~\cite{chaudhary2023code}, Evol-Instruct-Code~\cite{luo2023wizardcoder}, OSS-Instruct~\cite{wei2023magicoder}, PackageInstruct~\cite{huang2025opencoder}, and KODCODE~\cite{xu2025kodcode}. 
For code execution reasoning, existing works~\cite{ding2024semcoder,li2025codei,chen2025chain,jung2025code} typically passively mine or generate code from real-world code snippets. 
In contrast, \approach actively synthesizes data by applying structural constraints and difficulty filtering, resulting in higher-quality data (see Section~\ref{sec:data_synthesis}).

\section{Conclusion and Future Work}
\label{sec:conclusion_and_future_work}

In this work, we presented \approach, a post-training framework for teaching code LLMs to reason about program execution. On the data side, we build a constraint-based synthesis pipeline that actively generates executable programs under structural constraints, augments them with diverse inputs, and applies difficulty-aware filtering to form a curriculum-style dataset containing challenging yet solvable instances. On the learning side, we propose a two-stage training pipeline: Stage~I focuses on execution reasoning via white-box reinforcement learning, rewarding the model for answering verifiable questions about intermediate control flow and variable states, and Stage~II adapts the model to code generation with unit-test–based rewards. Experiments show that a 7B code model trained with \approach achieves performance competitive with 32B models on execution reasoning benchmarks, and demonstrates improvement over post-training baselines on standard code generation benchmarks.

In future work, we plan to extend \approach along three directions. 
First, on the data side, we will broaden the coverage of types and methods in our synthesis pipeline and include more libraries. 
Second, we aim to generalize our framework to other programming languages. 
Finally, we intend to move from function-level snippets to project-level code and model the execution process of multi-file and project-level programs.

\section{Limitations}
\label{sec:limitations}
Our synthesized data is currently restricted to Python and mainly covers built-in types with limited library usage, so real-world code coverage is incomplete. Our experiments are conducted at the function or snippet level rather than on multi-file or project-level code. In addition, the proposed white-box reinforcement learning pipeline incurs higher computational and engineering overhead than standard supervised fine-tuning.

\clearpage
\bibliography{custom}

\clearpage
\appendix
\label{sec:appendix}
\section{Dataset Statistics and Analysis}

\subsection{Difficulty Imbalance in Existing Execution Training Datasets}
\label{subsection:Appendix-Difficulty-Imbalance}
To better understand the existing execution-style training datasets, 
we conducted a small-scale empirical study on two widely used training datasets from SEMCODER~\cite{ding2024semcoder} and CODEIO~\cite{li2025codei}.

On a random sample of 15k test cases from \textsc{SEMCODER}, the
Qwen2.5-Coder-Instruct-7B model already solves roughly 70\% of problems in a single
{pass@1} attempt, without any additional reasoning-specific fine-tuning. 
This suggests that a large portion of {SEMCODER} is trivial for modern code LLMs and 
provides limited signal for improving execution reasoning.

In contrast, when we randomly sample 15k problems from the training dataset \textsc{CODEIO}, we observe the opposite
phenomenon: even the strong model qwq32b~\cite{team2024qwq} frequently fails to find any
solution. 
A significant subset of 52.1\% \textsc{CodeIO} instances is unsolved by 
current frontier models. Independent evidence from REASONING GYM~\cite{stojanovski2025reasoning} further supports this picture: in their dataset,
the {CodeIO} programs are explicitly configured as high-difficulty ``code'' tasks, and
the reported zero-shot accuracies of strong reasoning models like QwQ-32B on these tasks remain low even under the easy
settings.

\subsection{Difficulty and Complexity Distribution}
\label{subsection:Appendix-Difficulty-Complexity-Distribution}

\begin{figure*}[t]
  \centering
  \includegraphics[width=\linewidth]{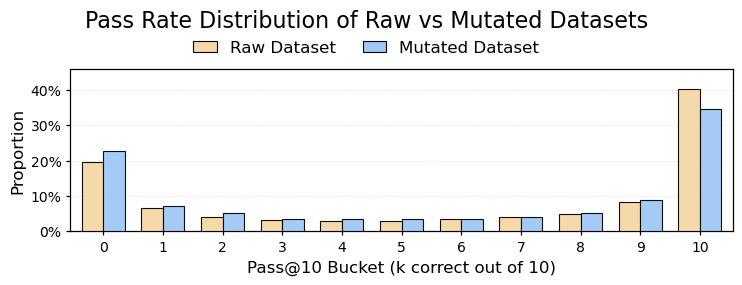}
  \caption{Difficulty distribution of the raw and mutated datasets, measured by the number of successful trials $k$ out of 10 for each synthesized problem.}
  \label{fig:difficulty_distribution}
\end{figure*}

Figure~\ref{fig:difficulty_distribution} shows the difficulty distribution of our synthesized problems, measured by the number of successful trials obtained by a baseline code model Qwen2.5-Coder-7B-Instruct on the raw and mutated datasets. For each problem, we run the model with temperature 1.0 with the task input-output prediction for ten independent trials and record the number of trials $k \in \{0,\dots,10\}$ whose predictions are correct. The histogram indicates that both datasets cover a wide range of difficulty levels, and that input mutation slightly shifts probability mass from trivially easy problems toward harder ones while preserving overall diversity.

Table~\ref{tab:code-complexity} reports several structural complexity metrics computed over the final difficulty-filtered dataset. On average, each snippet contains 9.93 non-empty, non-comment lines of code (LOC), with a median of 9. The maximum depth of the full Python abstract syntax tree (AST) has a mean of 9.74 and a median of 10, reflecting non-trivial expression structure even for relatively short snippets. Each snippet includes on average 1.43 branch constructs (e.g., \texttt{if}/\texttt{elif}/ternary expressions) and 0.85 loop constructs (e.g., \texttt{for}/\texttt{while} and comprehensions), both with medians of 1. 

To better characterize control flow, we additionally measure the nesting depth of structured blocks, counting only \texttt{if}, \texttt{for}, \texttt{while}, \texttt{try}, \texttt{with}, function definitions, and class definitions. The resulting control-flow nesting depth has a mean of 2.86 and a median of 3, indicating that most instances involve multiple layers of nested logic rather than flat scripts.

\begin{table}[t]
\centering
\small
\begin{tabular}{lrr}
\toprule
Metric & Mean & Median \\
\midrule
Lines of code (LOC)              & 9.93 & 9 \\
AST depth (full syntax tree)     & 9.74 & 10 \\
\#branches (if / elif / ternary) & 1.43 & 1 \\
\#loops (for / while / comp.)    & 0.85 & 1 \\
Control-flow nesting depth       & 2.86 & 3 \\
\bottomrule
\end{tabular}
\caption{Code complexity statistics of the final difficulty-filtered dataset. 
AST depth is measured as the maximum depth of the full Python abstract syntax tree, 
while control-flow nesting depth counts only structured blocks such as 
\texttt{if}/\texttt{for}/\texttt{while}/\texttt{try}/\texttt{with}/function and class definitions.}
\label{tab:code-complexity}
\end{table}

\subsection{Type Distribution}
\label{subsection:Appendix-Type-Distribution}
\begin{figure*}[t]
  \centering
  \includegraphics[width=\linewidth]{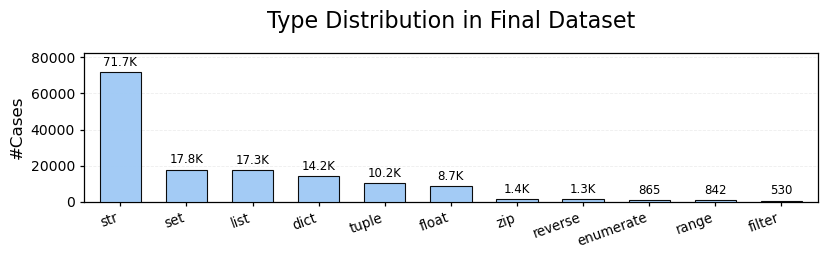}
  \caption{Distribution of Python built-in types in the final difficulty-filtered dataset.}
  \label{fig:type_distribution}
\end{figure*}
Figure~\ref{fig:type_distribution} presents the distribution of Python built-in types and related operations in the final difficulty-filtered dataset. String-manipulation problems (\texttt{str}) constitute nearly half of all instances, followed by sets (\texttt{set}), lists (\texttt{list}), and dictionaries (\texttt{dict}). There are also less frequent but still tasks such as tuples, floating-point numbers, and operations such as \texttt{zip}, \texttt{enumerate}, \texttt{reverse}, \texttt{range}, and \texttt{filter}. This mix of frequent and long-tail types ensures that the model is exposed to a broad spectrum of everyday Python programming primitives.

\subsection{Filtering Statistics}
\label{subsection:appendix-filtering-statistics}

\begin{figure}[t]
  \centering
  \includegraphics[width=\linewidth]{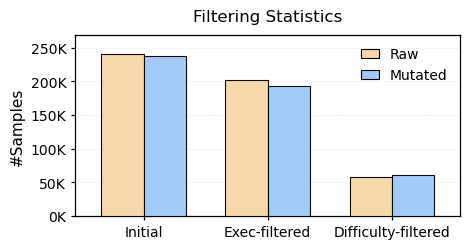}
  \caption{Filtering statistics of the synthesized raw and mutated datasets, showing the number of samples that remain after execution-based and difficulty-based filtering stages.}
  \label{fig:filtering_statistics}
\end{figure}

Figure~\ref{fig:filtering_statistics} summarizes the effect of our filtering stages on both the raw and mutated datasets. Starting from 239{,}992 raw samples and 239{,}466 mutated samples, execution-based filtering removes snippets that fail to run successfully or violate basic output constraints (e.g., runtime exceptions, timeouts, or excessively long outputs), leaving 201{,}537 raw and 191{,}463 mutated samples. We then apply difficulty-based filtering using the success-count distribution described in Section~\ref{subsec:Input-Synthesis-and-Data-Filtering}, retaining 119{,}358 instances that are non-trivial for the baseline model. The decrease across stages illustrates how each component of the pipeline progressively improves data quality while preserving a large and diverse training set.

\subsection{Contamination Analysis}
\label{subsection:Appendix-Contamination-Analysis}
 \begin{figure}[t]
  \centering
  \includegraphics[width=\linewidth]{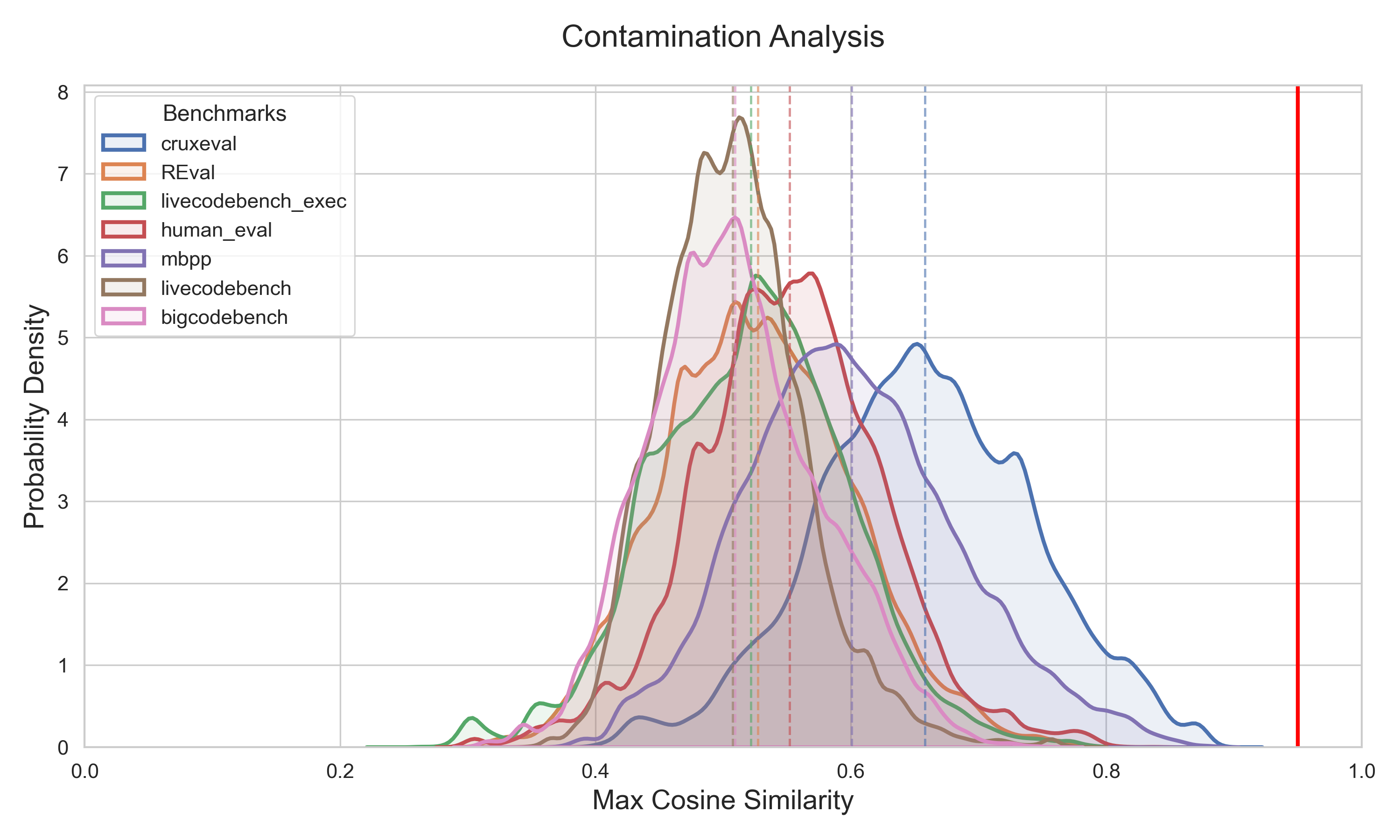}
  \caption{Contamination analysis between our synthesized dataset and downstream evaluation benchmarks, showing the distribution of maximum cosine similarity scores per training instance and the 0.95 threshold used to flag potential overlaps.}
  \label{fig:contamination_analysis}
\end{figure}
We evaluate potential contamination between our synthesized training data and existing evaluation benchmarks following the same embedding-based protocol as KODCODE. For each question in our dataset, we encode its natural-language description using the \texttt{all-mpnet-base-v2} sentence-embedding model, and apply the same encoder to all problems from our downstream benchmarks. We then compute cosine similarity between each training instance and all benchmark questions, and record the maximum similarity for each instance.
Following prior work, we adopt $0.95$ as a conservative similarity threshold for flagging potential contamination. Figure~\ref{fig:contamination_analysis} shows the distribution of maximum cosine similarity scores across all training instances, together with a vertical line at $0.95$. In our data, \textbf{no training instance exceeds this threshold}, and the entire distribution lies clearly below $0.95$, suggesting that near-duplicates or paraphrased copies of benchmark items are extremely rare.
We additionally perform a manual inspection of the few highest-similarity pairs (e.g., instances with similarity close to the right tail of the distribution) and confirm that they differ substantially in both surface form and semantics. Consequently, we do not remove any training examples at this stage and consider our synthesized dataset to be effectively contamination-free with respect to the benchmarks used in our evaluation.

\section{Data Synthesis Details}
This section provides the specific prompt templates and logical details used in the
\emph{Constraint-Based Data Synthesis} pipeline described in Section~\ref{sec:data_synthesis},
to support the reproducibility of our experiments.

\subsection{Code Synthesis Prompts with Constraints}
\label{app:code-synthesis-prompts}

To synthesize executable programs that exhibit non-trivial execution behavior, we use QWQ-32B as the generator model with explicit structural constraints.
Figure~\ref{fig:prompt_with_constraints} shows a representative prompt template used
for generating Python code that tests the \texttt{rstrip} method of the
\texttt{str} type. The system message positions the model as an expert Python
programmer and instructs it to strictly adhere to the given constraints.

The user prompt specifies: (i) \emph{control-structure constraints}, such as the
requirement to include a \texttt{for} loop and to nest an \texttt{if} statement
inside a \texttt{while} loop; (ii) \emph{method-call constraints}, such as invoking
the target method multiple times and combining it with at least one additional
built-in method; and (iii) formatting requirements, such as avoiding comments and
emitting a single Markdown code block. By enforcing such constraints at generation
time, we obtain programs that naturally contain nested control flow and rich
interactions among multiple built-in operations.

\paragraph{Curriculum levels.}
To encourage the model to gradually acquire more complex execution patterns, we organize
our constraint-based prompts into three curriculum levels based on the required control
flow and method interactions. Table~\ref{tab:curriculum-levels} summarizes the design.

% in preamble:
% \usepackage{tabularx}

\begin{table*}[t]
\centering
\small
\setlength{\tabcolsep}{4pt}        % 可视情况调到 3pt
\renewcommand{\arraystretch}{1.05} % 略微增大行高，更好读

\begin{tabularx}{\textwidth}{@{}c >{\raggedright\arraybackslash}X r@{}}
\toprule
Level & Structural constraints & \shortstack[r]{Share\\(\%)} \\
\midrule
1 & Single built-in method call without any control-flow constructs. & 20.9 \\
2 & Multiple method calls with at least one shallow \texttt{if} or \texttt{for} statement, but no nested control-flow blocks. & 14.2 \\
3 & Multiple method calls with nested \texttt{if}/\texttt{for}/\texttt{while} statements, and a maximum control-flow nesting depth of 3. & 64.9 \\
\bottomrule
\end{tabularx}

\caption{Curriculum levels used in constraint-based code synthesis, grouped by the
required control-flow structure and method interactions in the final dataset.}
\label{tab:curriculum-levels}
\end{table*}

\begin{figure*}[h]
\begin{promptbox}[An example of prompt with constraints]
% --- System Prompt 部分 ---
\textbf{System Prompt:} You are an expert Python programmer. Your task is to generate valid, executable Python code that strictly adheres to a complex set of structural and logical constraints.

\vspace{0.5em}
\noindent\rule{\linewidth}{0.4pt} % 添加一条水平分割线，区分 System 和 User
\vspace{0.5em}

% --- User Prompt 部分 ---
\textbf{User Prompt:}

\textbf{Task:} Write Python code that tests the \texttt{rstrip} method of the \texttt{str} type.

\vspace{0.5em}
\textbf{Constraints:}
\begin{itemize}[leftmargin=1.5em, nosep] % 使用紧凑列表
    \item \textbf{Control Structure Constraints:}
    \begin{itemize}[leftmargin=1.5em, nosep]
        \item The code must include a \texttt{for} statement.
        \item You must nest an \texttt{if} statement inside the \texttt{for} loop.
    \end{itemize}
    
    \vspace{0.3em} %稍微增加一点间距
    
    \item \textbf{Method Call Constraints:}
    \begin{itemize}[leftmargin=1.5em, nosep]
        \item Invoke the test method multiple times within the code.
        \item Integrate at least one additional built-in method (preferably from \texttt{str}, \texttt{list}, \texttt{set}, or \texttt{tuple}).
    \end{itemize}
\end{itemize}

\vspace{0.8em}

\textbf{Formatting:}
\begin{itemize}[leftmargin=1.5em, nosep]
    \item Do not include any comments.
    \item Call the entry-point function exactly once.
\end{itemize}

\vspace{0.8em}

\textbf{Output Format:} \\
You must output a markdown code block:

\verb|```|python\\
complete code snippet\\
\verb|```| \\

\end{promptbox} \caption{A constraint-based code synthesis prompt used for generating Python programs that test specific built-in methods while strictly adhering to structural and formatting requirements.} \label{fig:prompt_with_constraints} \end{figure*}

\subsection{Input Synthesis and Mutation}
\label{subsection:Appendix-Input-Synthesis-Mutation}
\begin{figure*}[h]
\begin{promptbox}[An example of prompt for input mutation]

% --- Task Definition ---
\textbf{Task:} Modify the arguments in the provided code based on specific constraints.

\vspace{0.5em}
\textbf{Provided Code:}
\begin{tcolorbox}[colback=white, colframe=gray!20, boxrule=0.5pt, sharp corners, left=2pt, top=2pt, bottom=2pt]
\small\ttfamily
def test\_rstrip(s):\\
\hspace*{1.5em} result = s.rstrip()\\
\hspace*{1.5em} for char in result:\\
\hspace*{3em} if char.isalpha():\\
\hspace*{4.5em} result = result.lstrip(char)\\
\hspace*{3em} elif char.isdigit():\\
\hspace*{4.5em} result = result.strip(char)\\
\hspace*{3em} else:\\
\hspace*{4.5em} result = result.rstrip(char)\\
\hspace*{1.5em} return result\\
assert test\_rstrip("  hello world  ") == "  hello world"
\end{tcolorbox}

\vspace{0.2em}

% --- Instructions ---
\textbf{Instructions:}
\begin{itemize}[leftmargin=1.5em, nosep]
    \item Reconstruct the arguments in the function call by \textbf{directly writing the arguments} inside the function call (do not assign them to variables first).
    \item You must use \textbf{one or more} values from the reference list below.
\end{itemize}

\vspace{0.5em}

% --- Reference Values ---
\textbf{Reference Values:}
\begin{itemize}[leftmargin=1.5em, nosep]
    \item \textbf{Integer Values:} \texttt{12, 7, 11, 9, 11, 7, 14}
    % 下面这行是报错的根源，所有的 % 都必须写成 \%
    \item \textbf{String Values:} \texttt{'EEFujAr'}, \texttt{'E2NC97aoEt'}, \texttt{'ZWRus3xdc8'}, \texttt{'a-s*?Rx\&;'}, \texttt{'STxJPmNuB4'}, \texttt{'TK07iWVF0'}, \texttt{'\%\%\%\%\%\%\%\%\%\%\%\%\%'}
\end{itemize}

\vspace{0.5em}

% --- Notes/Constraints ---
\textbf{Notes \& Constraints:}
\begin{itemize}[leftmargin=1.5em, nosep]
    \item For types such as \texttt{str}, \texttt{dict}, \texttt{list}, or \texttt{set}, try to increase their length.
    \item You may modify the code if needed, as long as it aligns with the reference values.
\end{itemize}

\vspace{0.8em}
\noindent\rule{\linewidth}{0.4pt} % 分割线
\vspace{0.5em}

% --- Output Format ---
\textbf{Output Format:} \\
Call the entry point function with the new arguments and use the print statement to print the result. Return only a single Markdown code block:
\begin{tcolorbox}[colback=white, colframe=gray!50, boxrule=0.5pt, sharp corners, width=0.6\linewidth]
\texttt{\textasciigrave\textasciigrave\textasciigrave python} \\
\# complete code snippet \\
\texttt{\textasciigrave\textasciigrave\textasciigrave}
\end{tcolorbox}

\end{promptbox}
\caption{An example of prompt for mutating inputs in the code}
\label{fig:mutate_prompt}
\end{figure*}

Given a synthesized program, we create diverse inputs to probe its
execution behavior. We first use the same QWQ-32B generator to construct an input. We then perform type-aware input mutation to obtain
valid samples with mutated inputs.

Figure~\ref{fig:mutate_prompt} presents the prompt used for input mutation.
The template exposes the original code snippet and asks the model to directly rewrite
the arguments of the entry-point function call, while respecting a list of reference
values (e.g., candidate integers and strings) and a set of mutation guidelines.
These guidelines instruct the model to, for example, increase the length of
strings or containers and to modify arguments in a way that remains consistent with
the reference values and program semantics. The output is again required to be a
single Markdown code block that calls the entry function and prints the result.

\begin{figure*}[h]
\begin{promptbox}[An example of Input Mutation]

% --- Task Definition ---

\vspace{0.5em}
\textbf{Code before input mutation:}
\begin{tcolorbox}[colback=white, colframe=gray!20, boxrule=0.5pt, sharp corners, left=2pt, top=2pt, bottom=2pt]
\small\ttfamily
def test\_rstrip(s):\\
\hspace*{1.5em} result = s.rstrip()\\
\hspace*{1.5em} for char in result:\\
\hspace*{3em} if char.isalpha():\\
\hspace*{4.5em} result = result.lstrip(char)\\
\hspace*{3em} elif char.isdigit():\\
\hspace*{4.5em} result = result.strip(char)\\
\hspace*{3em} else:\\
\hspace*{4.5em} result = result.rstrip(char)\\
\hspace*{1.5em} return result \\
assert test\_rstrip("  hello world  ") == "  hello world"
\end{tcolorbox}

\vspace{0.8em}
\noindent\rule{\linewidth}{0.4pt} % 分割线
\vspace{0.5em}

% --- Output Format ---
\vspace{0.5em}
\textbf{Code after input mutation:}
\begin{tcolorbox}[colback=white, colframe=gray!20, boxrule=0.5pt, sharp corners, left=2pt, top=2pt, bottom=2pt]
\small\ttfamily
def test\_rstrip(s):\\
\hspace*{1.5em} result = s.rstrip()\\
\hspace*{1.5em} for char in result:\\
\hspace*{3em} if char.isalpha():\\
\hspace*{4.5em} result = result.lstrip(char)\\
\hspace*{3em} elif char.isdigit():\\
\hspace*{4.5em} result = result.strip(char)\\
\hspace*{3em} else:\\
\hspace*{4.5em} result = result.rstrip(char)\\
\hspace*{1.5em} return result \\
% 这里使用了 \colorbox{颜色}{内容} 和 \textbf{内容} 来高亮这一行
% red!15 表示 15% 的红色混合 85% 的白色，形成淡红色背景
\colorbox{red!15}{\textbf{assert test\_rstrip('E2NC97aoEt') == ""}}
\end{tcolorbox}

\end{promptbox}
\caption{An example of input mutation applied to the synthesized code that tests the \texttt{rstrip} method. The mutated input is highlighted.}
\label{fig:input_mutation_example}
\end{figure*}

Figure~\ref{fig:input_mutation_example} shows an example of the resulting mutation.
The top part displays the original code together with its initial assertion,
while the bottom part shows the mutated version produced by our procedure.
The highlighted assertion demonstrates how the input string is replaced by a
synthetic value that changes the execution trace yet still exercises
the same functionality.

\subsection{White-Box Question Generation}
\label{appendix:white-box_question_generation}
\begin{figure*}[h]
\begin{promptbox}[An example of white-box questions]

\small\ttfamily
You are a programming expert.
Your task is to analyze the Python code and answer the questions by simulating the execution step by step.

\vspace{0.5em}

Here is the code content:
\begin{tcolorbox}[colback=white, colframe=gray!30, boxrule=0.5pt, sharp corners, left=2pt, top=2pt, bottom=2pt]
\small\ttfamily
1\hspace*{1.5em}def test\_rstrip(s):\\
2\hspace*{3em}result = s.rstrip()\\
3\hspace*{3em}for char in result:\\
4\hspace*{4.5em}if char.isalpha():\\
5\hspace*{6em}result = result.lstrip(char)\\
6\hspace*{4.5em}elif char.isdigit():\\
7\hspace*{6em}result = result.strip(char)\\
8\hspace*{4.5em}else:\\
9\hspace*{6em}result = result.rstrip(char)\\
10\hspace*{1.5em}return result\\
11\hspace*{1.5em}assert test\_rstrip("  hello world  ") == ????
\end{tcolorbox}

\vspace{0.5em}
Here are the questions:\\
\textbf{Question1:} Fill the assertion statement.\\
\textbf{Question2:} What is the value and type of the variable \texttt{result} after \textbf{Line 2} (\texttt{result = s.rstrip()}) is executed for the \textbf{1st time}?\\
\textbf{Question3:} Tracing the execution, which line is executed immediately after \textbf{Line 6} (\texttt{elif char.isdigit():}) is executed for the \textbf{1st time}?\\
\textbf{Question4:} Tracing the execution, which line is executed immediately after \textbf{Line 9} (\texttt{result = result.rstrip(char)}) is executed for the \textbf{1st time}?\\
\textbf{Question5:} What is the value and type of the variable \texttt{result} after \textbf{Line 5} (\texttt{result = result.lstrip(char)}) is executed for the \textbf{1st time}? \\
\textbf{Question6:} Tracing the execution, which line is executed immediately after \textbf{Line 4} (\texttt{if char.isalpha():}) is executed for the \textbf{3rd time}?

\vspace{0.8em}
\textbf{Guidelines for “next statement” questions:}
\begin{itemize}[leftmargin=1.5em, nosep]
    \item Determine the next line executed after the given statement.
    \item CRITICAL: Your answer MUST be the exact, verbatim source code of the next line — copied character-for-character, including indentation.
    \item Do NOT include line numbers, quotes, backticks, comments, or any extra words.
\end{itemize}

\vspace{0.5em}
\textbf{Guidelines for “type \& value” questions:}
\begin{itemize}[leftmargin=1.5em, nosep]
    \item STRICT FORMAT: \texttt{value; type} (Exactly one semicolon and one space).
    \item Example: \texttt{1; int} \quad \texttt{'hello'; str} \quad \texttt{[1, 2]; list}
\end{itemize}

\vspace{0.5em}
\textbf{ABSOLUTE FORMAT RULES (MUST FOLLOW):}
\begin{itemize}[leftmargin=1.5em, nosep]
    \item Output all answers one per line and in the listed order.
    \item For “next statement” answers: output ONLY the code statement string. Do not output line numbers!
\end{itemize}

\vspace{0.5em}
Format your response strictly as follows:\\
<reasoning>\\
your step-by-step reasoning here\\
</reasoning>\\
<answer>\\
Answer for question1\\
...\\
Answer for question5\\
</answer>

\end{promptbox}
\caption{A white-box question prompt used during reinforcement learning, combining a fully instrumented code snippet with multiple next-statement and value-and-type questions derived from its execution trace, together with strict formatting rules for the model’s reasoning and answers.}
\label{fig:white_box_question_example}
\end{figure*}

To obtain supervision that directly targets execution reasoning, we convert each
instrumented program run into a set of \emph{white-box questions} derived from its
execution trace. Concretely, we execute the synthesized Python code in a sandbox and
use the built-in \texttt{traceback} facility to record the sequence of executed
statements together with the evolving program state.

Given an execution trace, we construct two types of questions:

\begin{itemize}
    \item \textbf{Variable-state (data-flow) questions.}
    For each executed statement, we compare the values of all in-scope variables
    immediately before and after the statement. Whenever a variable changes, we
    create a question that asks for its value and Python type after the statement
    has executed. These questions encourage the model to track how data flows
    through the program.

    \item \textbf{Next-statement (control-flow) questions.}
    For each executed statement, we inspect the next statement in the trace.
    If the current statement is a control-flow construct such as \texttt{if},
    \texttt{while}, or \texttt{for}, we create a question asking for the exact
    source line that will be executed next. In addition, whenever the line number
    of the next executed statement is smaller than that of the current one
    (i.e., control transfers backwards in the source file, as in loop iterations
    or taken branches), we also generate a next-statement question. These
    questions require the model to reason about branch conditions and loop
    behavior.
\end{itemize}

All candidate questions from a trace are collected into a problem set for the
corresponding program. For each training instance, we shuffle this set and sample
up to ten questions. Some traces produce fewer than ten valid questions, so the
resulting number of white-box questions per instance varies. On average, each
program contributes 7.8 white-box questions, of which 3.2 are control-flow
(next-statement) questions and 4.6 are data-flow (variable-state) questions. This
yields dense supervision over both control-flow and value-tracking aspects of
execution.

Figure~\ref{fig:white_box_question_example} illustrates the prompt used to query the model
with these white-box questions. The upper part displays the full code snippet with
line numbers, while the lower part lists several next-statement and variable-state
questions derived from a single execution trace, together with strict formatting
rules for the model’s reasoning and answers.

\section{Experimental Setup}
\subsection{Supervised Fine-Tuning (SFT)}
\label{subsection: Appendix-Supervised-Fine-Tuning-(SFT)}
Table~\ref{tab:sft-hparams} summarizes the hyper-parameters used in the
supervised fine-tuning (SFT) stage.
We fine-tune {Qwen2.5-Coder-7B-Instruct} in a full-parameter
setting using the {LLaMAFactory}~\cite{zheng2024llamafactory} framework with DeepSpeed ZeRO-2
on a cluster of {8$\times$H100} GPUs.

For SFT, we first 
randomly sample 30K examples from the dataset constructed in section~\ref{sec:data_synthesis} to form the \texttt{sft\_new\_dataset}
split used for training.
We apply the official \texttt{qwen} chat template and truncate each
sequence to at most 4096 tokens.
Data preprocessing uses 16 CPU workers and data loading uses 4 workers.
Unless otherwise specified, all SFT experiments in this paper follow this configuration.

\begin{table}[t]
\centering
\small
\begin{tabular}{l c}
\toprule
\textbf{Hyper-parameter} & \textbf{Value} \\
\midrule
Learning rate & $1 \times 10^{-5}$ \\
Number of epochs & 2 \\
Per-device batch size & 2 \\
Gradient accumulation steps & 8 \\
Effective batch size & 128 \\
Max sequence length & 4096 \\
Optimizer & AdamW \\
Learning rate scheduler & cosine \\
Warmup ratio & 0.1 \\
Precision & bfloat16 \\
Fine-tuning type & full-parameter \\
Parallelism & DeepSpeed ZeRO-2 \\
\bottomrule
\end{tabular}
\caption{Hyper-parameters for supervised fine-tuning (SFT).}
\label{tab:sft-hparams}
\end{table}

\subsection{Reinforcement Learning (GRPO)}
\label{subsection: Reinforcement-Learning-(GRPO)}

Table~\ref{tab:grpo-hparams} summarizes the hyper-parameters used in the
GRPO-based reinforcement learning stages.

\paragraph{Stage I: reasoning RL.}
In the first RL stage, we start from the SFT checkpoint of
\textbf{Qwen2.5-Coder-7B-Instruct} and apply GRPO on our synthetic
execution-reasoning corpus.
The reward combines input--output correctness and white-box signals
(control-flow and data-flow questions derived from execution traces).

\paragraph{Stage II: code generation RL.}
In the second RL stage, we start from the Stage-I checkpoint and apply
GRPO with the \textsc{VeRL}~\cite{sheng2024hybridflow} framework on the PrimeCode
(\texttt{eurus\_prime}) train/validation splits.

Training is also performed on \textbf{8$\times$H100} GPUs with FSDP
(parameter and optimizer offloading) and gradient checkpointing enabled.

Unless otherwise specified, \textbf{both RL stages share the same GRPO
hyper-parameters} as listed in Table~\ref{tab:grpo-hparams}; only the
training data and reward functions differ between the two stages.

\begin{table}[t]
\centering
\small
\begin{tabular}{l c}
\toprule
\textbf{Hyper-parameter} & \textbf{Value} \\
\midrule
Algorithm & GRPO (VeRL) \\
Train batch size (prompts) & 128 \\
\#responses per prompt $n$ & 8 \\
Max prompt length & 2048 \\
Max response length & 4096 \\
Actor learning rate & $1 \times 10^{-6}$ \\
PPO mini-batch size & 64 \\
PPO micro-batch size / GPU & 1 \\
Rollout backend & vLLM \\
Tensor model parallel size & 4 \\
KL in loss / in reward & on / off (coef $0.0$) \\
Total steps & 500 \\
Parallelism & FSDP (param \& opt offload) \\
GPUs & 8$\times$H100 \\
\bottomrule
\end{tabular}
\caption{Hyper-parameters for GRPO-based reinforcement learning (UT-RL) on the PrimeCode dataset.}
\label{tab:grpo-hparams}
\end{table}

\section{Additional Ablations and Training Dynamics}

\subsection{Training Dynamics}
\label{appendix:training_dynamics}
\begin{figure}[t]
  \centering
  \includegraphics[width=\linewidth]{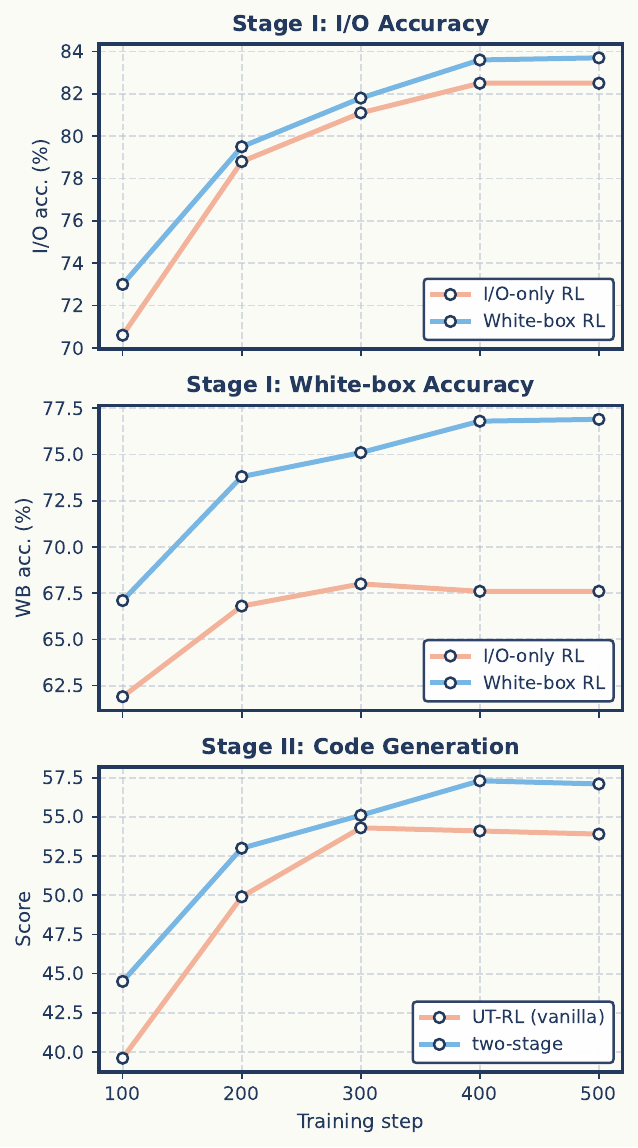}
  \caption{Stage I Reasoning and Stage II Code Generation: Training Curve Comparison.}
  \label{fig:training-dynamics}
\end{figure}

Figure~\ref{fig:training-dynamics} provides the training dynamics of Stage I and Stage II. In the Stage I curves, we observe that white-box RL yields stable gains on both white-box questions and I/O prediction. In the Stage II curve, we see that our two-stage training framework, which first trains the model for code reasoning and then trains it for code generation, consistently outperforms directly training the model for code generation from scratch, delivering stable improvements throughout training.

\subsection{Sensitivity to the Reward Mixing Coefficient $\alpha$}
\label{appendix：coefficient-sensitivity}
In Stage~I (white-box RL), we combine the I/O-based reward and the white-box reward via a convex mixture:
\begin{equation}
r = (1-\alpha)\, r_{\text{I/O}} + \alpha\, r_{\text{WB}},
\end{equation}
where $\alpha$ controls the relative weight of the white-box signal. In the main paper, we set $\alpha=0.5$.

We evaluate $\alpha \in \{0.25, 0.5, 0.75\}$ while keeping all other training settings fixed. Table~\ref{tab:alpha_sensitivity} reports CXEval-O, CXEval-I, LCB-O, and the fine-grained REval metrics (Coverage/State/Path/Output), along with the overall summary score. The experimental results indicate that performance is not sensitive to $\alpha$ within this range.

\begin{table*}[t]
\centering
\small
\setlength{\tabcolsep}{4.5pt}
\begin{tabular}{c|ccc|cccc|c}
\toprule
$\alpha$ & CXEval-O & CXEval-I & LCB-O & Coverage & State & Path & Output & Avg \\
\midrule
0.25 & 85.9 & 80.7 & 82.3 & 85.5 & 74.2 & 73.1 & 83.3 & 80.71 \\
0.50 & 85.6 & 81.0 & 82.3 & 85.8 & 74.5 & 73.0 & 83.2 & 80.77 \\
0.75 & 85.6 & 80.8 & 82.1 & 86.0 & 75.2 & 73.4 & 81.1 & 80.60 \\
\bottomrule
\end{tabular}
\caption{Sensitivity analysis of the reward mixing coefficient $\alpha$ in Stage~I. Higher is better for all metrics. Avg is the mean over the seven reported metrics, consistent with the main paper.}
\label{tab:alpha_sensitivity}
\end{table*}

\section{Experimental Details}
\subsection{Variant setup for Table~\ref{tab:code_reasoning}} \label{app:variant-setup}
All 7B variants in Table~\ref{tab:code_reasoning} are trained on the same pool of 60K programs
randomly sampled from our synthesized reasoning corpus.
Among them, 30K examples (15K I$\rightarrow$O and 15K O$\rightarrow$I) are used for the optional
SFT step.
The ``+ I/O O/I RL'' variant performs RL on all 60K examples (30K I$\rightarrow$O and 30K
O$\rightarrow$I) without SFT.
The ``+ SFT + I/O O/I RL'' variant uses the 30K split for SFT and runs RL on the remaining 30K
examples (15K I$\rightarrow$O and 15K O$\rightarrow$I).
The ``+ SFT + white-box RL'' variant shares the same 30K SFT split and performs RL on the
remaining 30K examples (15K white-box I$\rightarrow$O and 15K O$\rightarrow$I).

\subsection{Data-quality comparison setup}
\label{app:data-quality-setup}
To isolate the impact of data quality, we compare our synthesized data against three representative datasets: PYX-Sub~\cite{ding2024semcoder}, CodeI/O-Sub~\cite{li2025codei}, and Grounded-CoT~\cite{jung2025code}, which correspond to the two paradigms in Section~\ref{sec:introduction} (I/O CoT supervision vs.\ LLM-translated execution traces). For a fair comparison under the same budget, we sample matched-size training sets of 15K examples from each dataset and use a unified teacher, QwQ-32B~\cite{team2024qwq}, to generate all CoT and trace translations with the same prompting. We then fine-tune the same Qwen2.5-Coder-Instruct model on the I$\rightarrow$O prediction task using each 15K subset and report the results in Figure~\ref{fig:data-efficiency-scaling}.

\subsection{Ablation setups for data synthesis components}
\label{app:data-ablation-setup}
We conduct three SFT-only ablations, each using 15K training examples and the same fine-tuning protocol, evaluated on CRUXEval-O and LiveCodeBench-Exec. 
\textbf{Generating prompts with structural constraints:} \emph{Full-constraint} follows Section~\ref{sec:data_synthesis} by prompting the generator with specified types/methods and explicit structural constraints, while \emph{Weak-constraint} uses the same types/methods but removes structural constraints from the prompt. 
\textbf{Input synthesis:} using the same 15K code snippets, \emph{Full-input} includes multiple input configurations (original inputs and type-aware mutations) and samples to 15K examples, whereas \emph{Simple-input} keeps only one basic input per snippet. 
\textbf{Filtering by difficulty:} from the pool after execution filtering, \emph{Filtered} applies difficulty-aware filtering before uniformly sampling 15K examples, while \emph{No-filter} samples 15K examples directly without difficulty filtering.

\subsection{Construction of a Library-Involved I/O Prediction Benchmark}
\label{app:library-involved-i/o-prediction-benchmark-construction}

\begin{table*}[h]
\centering
\small
\setlength{\tabcolsep}{5pt}
\renewcommand{\arraystretch}{1.15}
\begin{tabularx}{\linewidth}{@{}lX@{}}
\toprule
\textbf{Category} & \textbf{Filtered libraries/APIs (keyword-level matching)} \\
\midrule
Randomness / stochasticity &
\code{random} (\code{shuffle}, \code{randint}, \code{choice(s)}, \code{seed}); \code{numpy.random}/\code{np.random}; \code{torch.rand}/\code{torch.randn}; \code{secrets}. \\

File I/O and filesystem traversal &
\code{open}/\code{with open}, \code{.read}, \code{.write}; \code{fileinput}; \code{pathlib}; \code{os.path}; \code{os} ops (\code{listdir}, \code{walk}, \code{scandir}, \code{remove}, \code{unlink}, \code{rmdir}, \code{mkdir}, \code{makedirs}, \code{rename}, \code{replace}, \code{stat}, \code{chmod}, \code{chown}, \code{getcwd}, \code{chdir}); \code{shutil}; \code{glob}; \code{tempfile}. \\

Archives and compression &
\code{zipfile}; \code{tarfile}; \code{gzip}; \code{bz2}; \code{lzma}. \\

Structured file readers/writers and external formats &
\code{csv} (\code{reader}, \code{dictreader}, \code{writer}, \code{dictwriter});
\code{pandas}/\code{pd} (\code{read\_csv}, \code{read\_table}, \code{read\_excel}, \code{read\_parquet}, \code{read\_feather}, \code{read\_json}, \code{read\_pickle});
writers (\code{to\_csv}, \code{to\_excel}); \code{openpyxl}. \\

Serialization / persistence &
\code{numpy}/\code{np} (\code{load}, \code{save}, \code{savez}, \code{savez\_compressed});
\code{torch} (\code{save}, \code{load});
\code{pickle} (\code{load}, \code{dump});
\code{joblib} (\code{load}, \code{dump});
\code{json.load};
\code{yaml} (\code{load}, \code{safe\_load}). \\

Databases &
\code{sqlite3.connect}. \\
\bottomrule
\end{tabularx}
\caption{Blacklist used to exclude non-deterministic or environment-dependent BigCodeBench solutions when constructing the library-involved I/O prediction set.}
\label{tab:libio-blacklist}
\end{table*}

\begin{figure*}[t]
\centering
\small
\begin{promptbox}[An example of constructing a library-involved I/O prediction benchmark]

% -------------------- Row 1: Complete prompt (full width) --------------------
\textbf{Complete prompt (from \texttt{complete\_prompt}):}
\begin{tcolorbox}[colback=white, colframe=gray!20, boxrule=0.5pt, sharp corners, left=3pt, top=3pt, bottom=3pt]
\scriptsize\ttfamily
from collections import Counter\\
import itertools\\[2pt]
def task\_func(d):\\
\hspace*{1.5em}"""\\
\hspace*{1.5em}Count the occurrence of each integer in the values of the input dictionary,\\
\hspace*{1.5em}where each value is a list of integers, and return a dictionary with these counts.\\[2pt]
\hspace*{1.5em}Requirements:\\
\hspace*{1.5em}- collections.Counter\\
\hspace*{1.5em}- itertools\\[2pt]
\hspace*{1.5em}Example:\\
\hspace*{1.5em}>>> d = \{'a': [1, 2, 3, 1], 'b': [3, 4, 5], 'c': [1, 2]\}\\
\hspace*{1.5em}>>> count\_dict = task\_func(d)\\
\hspace*{1.5em}>>> print(count\_dict)\\
\hspace*{1.5em}\{1: 3, 2: 2, 3: 2, 4: 1, 5: 1\}\\
\hspace*{1.5em}"""
\end{tcolorbox}

\vspace{0.55em}
\noindent\rule{\linewidth}{0.4pt}
\vspace{0.55em}

% -------------------- Row 2: Extracted I/O + Canonical solution (two columns) --------------------
\begin{minipage}[t]{0.48\linewidth}
\textbf{Extracted I/O (from the example):}
\begin{tcolorbox}[colback=white, colframe=gray!20, boxrule=0.5pt, sharp corners, left=3pt, top=3pt, bottom=3pt]
\scriptsize\ttfamily
d = \{'a': [1, 2, 3, 1], 'b': [3, 4, 5], 'c': [1, 2]\}\\
count\_dict = task\_func(d)\\
print(count\_dict)\\[2pt]
\textbf{stdout:} \{1: 3, 2: 2, 3: 2, 4: 1, 5: 1\}
\end{tcolorbox}
\end{minipage}
\hfill
\begin{minipage}[t]{0.48\linewidth}
\textbf{Canonical solution}
\begin{tcolorbox}[colback=white, colframe=gray!20, boxrule=0.5pt, sharp corners, left=3pt, top=3pt, bottom=3pt]
\scriptsize\ttfamily
from collections import Counter\\
import itertools\\[2pt]
def task\_func(d):\\
\hspace*{1.5em}count\_dict = Counter(\\
\hspace*{3em}itertools.chain.from\_iterable(d.values())\\
\hspace*{1.5em})\\
\hspace*{1.5em}return dict(count\_dict)
\end{tcolorbox}
\end{minipage}

\vspace{0.55em}
\noindent\rule{\linewidth}{0.4pt}
\vspace{0.55em}

% -------------------- Row 3: Final I/O prediction problem (full width) --------------------
\textbf{Final I/O prediction problem (model input):}
\begin{tcolorbox}[colback=white, colframe=gray!20, boxrule=0.5pt, sharp corners, left=3pt, top=3pt, bottom=3pt]
\scriptsize\ttfamily
\textbf{Your task is to fill the assert statement.}\\[4pt]
from collections import Counter\\
import itertools\\[2pt]
def task\_func(d):\\
\hspace*{1.5em}count\_dict = Counter(itertools.chain.from\_iterable(d.values()))\\
\hspace*{1.5em}return dict(count\_dict)\\[4pt]
d = \{'a': [1, 2, 3, 1], 'b': [3, 4, 5], 'c': [1, 2]\}\\
count\_dict = task\_func(d) \\
\colorbox{red!15}{\textbf{assert count\_dict == \textbf????}}
\end{tcolorbox}

\end{promptbox}
\caption{An example of constructing the library-involved I/O prediction benchmark from BigCodeBench. We show the original \texttt{complete\_prompt}, the extracted I/O pair, the executable \texttt{canonical\_solution}, and the final I/O prediction problem.}
\label{fig:libio_construction_example}
\end{figure*}

To evaluate transfer to library-dependent code, we construct a library-involved I/O prediction benchmark based on BigCodeBench. For each task, we extract an input–output pair from the example block in the task description (\texttt{complete\_prompt}) and treat the task’s (\texttt{canonical\_solution}) as the executable reference program.

Concretely, we recover the \emph{input} as the statements that define arguments and invoke the target function, and the \emph{output} as the printed result shown in the example. At evaluation time, we provide the extracted input and ask the model to predict the stdout output produced by executing the canonical solution. We report exact match accuracy.

To ensure determinism and avoid environment-specific behavior, we filter out tasks involving randomness or external resources. This is done using keyword-level matching over both the prompt and the solution, targeting stochastic APIs (e.g., random) and file I/O (e.g., open, pathlib, pickle). The full blacklist is provided in Table~\ref{tab:libio-blacklist}. After filtering, we obtain 241 test cases for evaluation. A detailed example is shown in Figure~\ref{fig:libio_construction_example}.

\section{Qualitative Analysis: The Impact of Code Reasoning}

\subsection{Improved Fine-grained Code Execution Understanding}

We conduct a qualitative analysis to demonstrate \approach's superior capability in tracing concrete execution steps compared to baselines. As illustrated in Figure~\ref{fig:case-study-io-vs-whitebox}, models trained solely with I/O O/I RL often struggle with control-flow logic, failing to evaluate guard conditions (e.g., if idx != idx2) and consequently "hallucinating" execution steps that do not occur. In contrast, \approach, trained via White-box RL, correctly interprets conditional statements and skips invalid loop iterations. This confirms that the model faithfully tracks intermediate variable states and adheres to the strict program logic.

\begin{figure*}[t]
\centering

% 定义颜色
\definecolor{boxheader}{RGB}{60, 60, 60}
\definecolor{boxbg}{RGB}{245, 245, 245}
\definecolor{codebg}{RGB}{255, 255, 255}

% 开始主盒子
\begin{tcolorbox}[
    enhanced,
    title=\textbf{Case Study: White-box RL vs. I/O O/I RL},
    coltitle=white,
    colbacktitle=boxheader,
    colframe=boxheader,
    colback=boxbg,
    fonttitle=\large\sffamily,
    boxrule=0.8pt,
    arc=0mm,
    left=6pt, right=6pt, top=6pt, bottom=6pt,
    drop fuzzy shadow, 
]

    % --- Question 部分 ---
    \textbf{Question.}
    \vspace{0.3em}
    
    % 代码块
    \begin{tcolorbox}[
        colback=codebg, 
        boxrule=0.5pt, 
        colframe=gray!30, 
        arc=0mm, 
        left=2pt, right=2pt, top=2pt, bottom=2pt
    ]
\begin{lstlisting}[
    language=Python,
    basicstyle=\ttfamily\small,
    numbers=left,
    numberstyle=\tiny\color{gray},
    numbersep=8pt,
    xleftmargin=1.5em,
    frame=none,
    aboveskip=0pt,
    belowskip=0pt
]
from typing import List, Tuple

def find_closest_elements(numbers: List[float]) -> Tuple[float, float]:
    closest_pair = None
    distance = None
    for idx, elem in enumerate(numbers):
        for idx2, elem2 in enumerate(numbers):
            if idx != idx2:
                if distance is None:
                    distance = abs(elem - elem2)
                    closest_pair = tuple(sorted([elem, elem2])) # Line 11
                else:
                    new_distance = abs(elem - elem2)
                    if new_distance < distance:
                        distance = new_distance
                        closest_pair = tuple(sorted([elem, elem2]))
    return closest_pair

# assert find_closest_elements([1.0, 2.0, 3.9, 4.0, 5.0, 2.2]) == ????
\end{lstlisting}
    \end{tcolorbox}

    \vspace{0.5em}
    \noindent Trace the execution of \texttt{find\_closest\_elements} with the given input. 
    Is Line 11 (\texttt{closest\_pair = ...}) executed? 
    If yes, what is the value and type of \texttt{closest\_pair} immediately after its \textbf{first execution}? 
    If the line is unreachable, answer \texttt{Nil}.
    
    \vspace{1em}
    \hrule
    \vspace{1em}

    % --- 对比部分 ---
    \begin{minipage}[t]{0.48\textwidth}
        {\large \textbf{I/O O/I RL (Incorrect)}} \par
        \vspace{0.5em}
        \textbf{Response.} I trace the loops starting with \texttt{idx=0} (\texttt{elem=1.0}). Inside the inner loop, I start with \texttt{idx2=0} (\texttt{elem2=1.0}). At this point, \texttt{distance} is \texttt{None}, so the code executes Line 11: \texttt{closest\_pair = tuple(sorted([1.0, 1.0]))}. Thus, the value is \texttt{(1.0, 1.0)} and the type is \texttt{tuple}. \par
        \vspace{0.5em}
        \textbf{Explanation.} The model \textbf{fails to evaluate the guard condition} \texttt{if idx != idx2}. It erroneously enters the conditional block for the self-pair indices $(0, 0)$, hallucinating an execution of Line 11 that actually never happens in the program flow.
    \end{minipage}%
    \hfill
    \vrule width 0.5pt
    \hfill
    \begin{minipage}[t]{0.48\textwidth}
        {\large \textbf{White-box RL (Correct)}} \par
        \vspace{0.5em}
        \textbf{Response.} I start with \texttt{idx=0}. For the inner loop, the first iteration is \texttt{idx2=0}. Since \texttt{idx == idx2}, the condition \texttt{if idx != idx2} is false, so I skip the body. The loop continues to \texttt{idx2=1} (\texttt{elem2=2.0}). Now the condition holds. Since \texttt{distance} is \texttt{None}, Line 11 executes: \texttt{closest\_pair = tuple(sorted([1.0, 2.0]))}. The value is \texttt{(1.0, 2.0)}. \par
        \vspace{0.5em}
        \textbf{Explanation.} The model \textbf{correctly traces the control flow}. It recognizes that the condition \texttt{idx != idx2} prevents execution for the first pair $(0,0)$, and correctly identifies the pair $(0,1)$ as the first valid execution point.
    \end{minipage}

\end{tcolorbox}

\caption{Case study comparing execution tracing capabilities. The \textbf{I/O O/I RL} model overlooks the conditional statement (Line 8). In contrast, the \textbf{White-box RL} model strictly follows the program logic, correctly skipping the first iteration and identifying the true first assignment.}
\label{fig:case-study-io-vs-whitebox}
\end{figure*}

\subsection{Benefit for Downstream Code Generation}

We further investigate how the fine-grained execution reasoning capability transfers to downstream code generation tasks. Figure~\ref{fig:case-study-generation} presents a qualitative comparison on the problem "Count Substrings With K-Frequency Characters I," which requires identifying substrings where at least one character meets a frequency threshold. The baseline I/O O/I RL model generates syntactically correct code but contains a logical error, confusing the required condition ("at least one") with a universal one ("for all"). It incorrectly enforces a stricter constraint (if 0 < f < k: valid = False), rejecting valid substrings. In contrast, \approach, enhanced by white-box reinforcement learning, correctly interprets the requirement and implements the logic using any(). This demonstrates that the fine-grained understanding of control flow and variable states acquired during reasoning training enables the model to perform a more accurate simulation of the program, leading to more robust handling of subtle logical constraints in code generation.

\begin{figure*}[t]
\centering

% --- 颜色定义 ---
\definecolor{boxheader}{RGB}{60, 60, 60}   % 深灰色标题栏
\definecolor{boxbg}{RGB}{245, 245, 245}    % 浅灰色背景
\definecolor{codebg}{RGB}{255, 255, 255}   % 代码块白色背景

% --- 主容器 ---
\begin{tcolorbox}[
    enhanced,
    title=\textbf{Case Study: Impact on Code Generation},
    coltitle=white,
    colbacktitle=boxheader,
    colframe=boxheader,
    colback=boxbg,
    fonttitle=\large\sffamily,
    boxrule=0.8pt,
    arc=0mm,
    left=6pt, right=6pt, top=6pt, bottom=6pt,
    drop fuzzy shadow, 
]

    % --- Question 部分 ---
    \textbf{Problem: Count Substrings With K-Frequency Characters I}
    \vspace{0.3em}
    
    % 问题描述框
    \begin{tcolorbox}[
        colback=codebg, 
        boxrule=0.5pt, 
        colframe=gray!30, 
        arc=0mm, 
        left=4pt, right=4pt, top=4pt, bottom=4pt
    ]
        \small
        Given a string \texttt{s} and an integer \texttt{k}, return the total number of substrings of \texttt{s} where \textbf{at least one character} appears \textbf{at least k times}.
        \par\vspace{0.3em}
        \textit{Example:} Input: \texttt{s = "abacb", k = 2}. Output: \texttt{4} ("aba", "abac", "abacb", "bacb").
    \end{tcolorbox}

    \vspace{1em}
    \hrule
    \vspace{1em}

    % --- 对比部分 (左右两栏) ---
    \begin{minipage}[t]{0.48\textwidth}
        {\large \textbf{I/O O/I RL (Incorrect)}} \par
        \vspace{0.5em}
        
        % 完整的错误代码
        \begin{tcolorbox}[
            colback=codebg, 
            frame hidden,      % <--- 修复报错：使用 frame hidden 而不是 frame=none
            boxrule=0pt,       % <--- 双重保险：边框宽度设为 0
            arc=0mm,
            left=2pt, right=2pt, top=2pt, bottom=2pt
        ]
\begin{lstlisting}[language=Python, basicstyle=\ttfamily\tiny, numbers=none, aboveskip=0pt, belowskip=0pt, breaklines=true]
class Solution:
    # Manually wrapped arguments to prevent ":" orphan
    def numberOfSubstrings(self, s: str, 
                           k: int) -> int:
        n = len(s)
        count = 0
        
        for i in range(n):
            freq = [0] * 26
            for j in range(i, n):
                char_index = ord(s[j]) - ord('a')
                freq[char_index] += 1
                
                valid = True
                for f in freq:
                    # INCORRECT LOGIC
                    if 0 < f < k:
                        valid = False
                        break
                if valid:
                    count += 1
        return count
\end{lstlisting}
        \end{tcolorbox}
        
        \vspace{0.3em}
        \textbf{Response Analysis.} The model generates a full solution but fails at the core logic. It correctly sets up the loops and frequency counting, but the validation condition \texttt{if 0 < f < k: valid = False} enforces that \textit{all} present characters must have frequency $\ge k$, violating the "at least one" requirement.
    \end{minipage}%
    \hfill
    \vrule width 0.5pt
    \hfill
    \begin{minipage}[t]{0.48\textwidth}
        {\large \textbf{\approach (Correct)}} \par
        \vspace{0.5em}

        % 完整的正确代码
        \begin{tcolorbox}[
            colback=codebg, 
            frame hidden,      % <--- 修复报错
            boxrule=0pt,       % <--- 修复报错
            arc=0mm,
            left=2pt, right=2pt, top=2pt, bottom=2pt
        ]
\begin{lstlisting}[language=Python, basicstyle=\ttfamily\tiny, numbers=none, aboveskip=0pt, belowskip=0pt, breaklines=true]
class Solution:
    # Manually wrapped arguments for clean layout
    def numberOfSubstrings(self, s: str, 
                           k: int) -> int:
        n = len(s)
        count = 0
        
        for i in range(n):
            char_count = [0] * 26
            for j in range(i, n):
                char_count[ord(s[j]) - ord('a')] += 1
                # CORRECT LOGIC
                valid = any(c >= k for c in char_count)
                if valid:
                    count += 1
        return count
\end{lstlisting}
        \end{tcolorbox}

        \vspace{0.3em}
        \textbf{Response Analysis.} The model correctly identifies the condition. By tracing the execution state during training (White-box RL), it understands that the validation should pass as soon as \textit{any} character meets the threshold, using \texttt{any(c >= k ...)} to correctly implement the logic.
    \end{minipage}

\end{tcolorbox}

\caption{Case study on code generation. The \textbf{I/O O/I RL} model produces syntactically correct code but fails logically by confusing "at least one" with "for all". In contrast, \textbf{\approach} (Ours), trained via white-box reinforcement learning, correctly implements the semantic requirement using \texttt{any()}. This demonstrates that grounding the model in fine-grained runtime behavior equips it with a deeper understanding of program semantics, enabling it to accurately handle subtle logical constraints that surface-level I/O training often misses.}
\label{fig:case-study-generation}
\end{figure*}

\section{Potential Risks}
\approach improves LLMs’ ability to reason about and generate executable code. As with other code LLM advances, this capability could be misused to produce harmful scripts or assist vulnerability exploitation. Moreover, generated code may still be incorrect or insecure. Therefore, all outputs should be treated as assistive suggestions and validated via human review before deployment.

\end{document}